\documentclass[twocolumn]{aastex62}

\usepackage{natbib}
\usepackage{CJKutf8}

\begin{document}
\begin{CJK}{UTF8}{bsmi}

\title{Large Scale Dynamo in a Primordial Accretion Flow -- An Interpretation from Hydrodynamic Simulation}

\author{Wei-Ting Liao}
\affil{Department of Astronomy, University of Illinois, 1002 West Green Street, Urbana, IL 61801, USA}
\email{wliao10@illinois.edu}

\author{Matthew Turk}
\affil{School of Information Sciences, University of Illinois, 501 East Daniel Street, Champaign, IL 61820, USA}
\affil{Department of Astronomy, University of Illinois, 1002 West Green Street, Urbana, IL 61801, USA}

\author{Hsi-Yu Schive}
\affil{Institute of Astrophysics, National Taiwan University, No.1 Sec.4 Roosevelt Road Taipei 10617, Taiwan}
\affil{Center for Theoretical Physics, National Taiwan University, No.1 Sec.4 Roosevelt Road Taipei 10617, Taiwan}

\begin{abstract}
Without an existing large scale coherent magnetic field in the early Universe, Population III (PopIII) stars would likely rotate at or near break-up speed. In this work, focusing on the accretion phase of PopIII stars, we investigate the possibility of generating a coherent magnetic field through large scale dynamo processes, as well as the corresponding field saturation level.
Using results from hydrodynamic simulations, we demonstrate that primordial accretion disks
are turbulent with a Shakura-Sunyaev disk parameter $\alpha_{ss} \gtrsim 10^{-3}$, and evidence helical turbulence with a dynamo number $\vert D_{\alpha \Omega} \vert \gg 10$. The presence of helical turbulence at these levels allows large scale dynamo modes to grow, and the saturation level is determined by the amount of net helicity remaining in the dynamo-active regions (a.k.a. the quenching problem). 
We demonstrate that, if the accretion could successfully alleviate the quenching problem, the magnetic field can reach approximate equipartition with $B/B_{\rm eq} \sim 3$.
\end{abstract}


\section{Introduction}
Although the formation of PopIII stars is one of the rarely well posed problems in Astronomy, with a properly understood initial condition from the cosmic microwave background measurement and a relative simple chemistry network that could be followed numerically, the currently unknown magnetic field in the early Universe presents itself as one of the biggest uncertainties. 

Magnetic fields in the present day star forming regions are known to play the major role in regulating the star formation processes, including braking the rotation of stars through magnetic braking and/or through star-disk interactions \citep{Mouschovias+91,Matt+05,Matt+10}. It could also cause the accretion disk around the protostar to be turbulent through magneto-rotational instability (MRI) \citep{Velikhov+59,Chandrasekhar+60,mri}.
Additionally, the presence of magnetic field is known to suppress the fragmentation \citep{Hennebelle+08,Price+08}, a coherent strong magnetic field could launch collimated jets \citep{LeeCF+18}, and, in some conditions, magnetic field could even forbid the disk formation -- the braking catastrophe \citep{Mellon+08,Mellon+09,LiZY+11}. 

In the early Universe, magnetic field is largely unknown, and is difficult, if not impossible, to measure. 
Theoretically, several magnetogenesis models have been proposed, yet it is hard to produce a strong and large scale coherent field \citep{Biermann+50,Weibel+59,Harrison+70}. Consequently, the role of magnetic fields in PopIII star formation processes falls back to the turbulent dynamo problem, i.e., the amplification of magnetic field through turbulence \citep{Tan+04_dynamo,Schleicher+10,Schober+12b,Turk+12,Latif+13,Schober+15}.

Magnetic field impacts PopIII stars in two major ways -- the fragmentation processes (and thus the mass and the initial mass function, IMF), and the rotataion of PopIII stars. Both of the mass and the rotation have major roles in affecting the stellar evolution path, and thus the subsequent chemical enrichment processes and the observables that imprinted on the immediate descendants -- low mass extremely metal poor (EMP) stars \citep{Maeder+12,Yoon+12}.

Several efforts, from both observational constraint and the results of numerical simulations, have been devoted to understand the mass and IMF of PopIII stars (observation: \citep{Aoki+14}, \cite{Ishigaki+18}; simulation: \cite{Clark+11}, \cite{Greif+11a}, \cite{Hosokawa+11}, \cite{Hirano+14}, \cite{Susa+14}, \cite{Stacy+16}, \cite{Hosokawa+16}). 
Although fragmentation is a common product in the PopIII star forming environments in numerical simulations and could potential build up the IMF for PopIII stars, currently only few simulations have explored the role of magnetic field \citep{Turk+12,Latif+13}, which is partly because the current computer power is not yet possible to fully resolve the dynamo processes and to investigate the saturation level.
Observationally, the peak of IMF is believed to be at $25 M_\odot$, which is determined through the chemical abundance of EMP stars, although the full shape of PopIII stars' IMF is still unclear. The unclear IMF at both the low and high mass end is due to the fact that the chemical abundance from these two mass ranges might not be imprinted in the EMP stars.

The other important component, the rotation of PopIII stars, is relatively unexplored \citep{Stacy+11_rotation,Choplin+19}. Rotation at the end of the pre-main sequence phase plays an important role in determining the stellar evolution path, since it governs the mass loss rate and the internal mixing \citep{Maeder+12}.
In present day star formation, the rotation of a star is in part regulated by magnetic field, although gravitational torque would also be important for massive stars \citep{LinMK+11, Rosen+12}.
Without an efficient braking mechanism, the star would gain angular momentum through accretion and ultimately reach nearly its break-up speed. 
Efficient magnetic braking requires a large scale, coherent magnetic field. In addition, the efficiency of magnetic braking depends on the strength of the magnetic field. As illustrated in \cite{Hirano+18}, a nearly equipartition large scale field is preferred for braking PopIII stars.

In the small-scale dynamo process, the growth of magnetic field is mediated through energy exchange with local turbulent kinetic energy. The small-scale dynamo proceeds quickly, with a rate comparable to the eddy turn overtime at the dissipation scale, and is able to reach a saturation level of $\mathcal{O}(1 \%) - \mathcal{O} (10 \%) $ compared to the kinetic turbulent energy \citep{Schober+15}. However, the produced field geometry is like to be entangled \citep{Schekochihin+04_morphology}.
The large scale dynamo, on the other hand, is able to produce a magnetic field beyond the turbulence driving scale with a super-equipartition saturation level \citep{Brandenburg+01}. 
An efficient large scale dynamo process would be a good candidate for braking the rotation of PopIII stars, although its efficiency is known to be limited by the subsequently developed magnetic current helicity.

Although magnetic field might have an interesting role in the PopIII star forming regions, current numerical simulations are not able to fully resolve the dynamo processes and thus are not able to study the role of magnetic field ab initio. The potential magnetic field geometry and magnitude in the early Universe thus relies on our understandings of dynamo processes from analytic modelings and from the turbulent statistics with a guide using results of turbulent box simulations. 
In this paper, using the results from pure hydrodynamic simulations, we aim to understand the possibility for a large scale dynamo to operate in a primoridal accretion flow, and to estimate the corresponding saturation level. 
We outline the numerical setup in S\ref{S:setup}, discuss the large scale dynamo process in S\ref{S:dynamo}, and discuss the results in S\ref{S: conclusion}.

\section{Simulation Setup}
\label{S:setup}
The simulation is carried out using GAMER \citep{gamer}, which we have augmented with a cylindrical fluid solver and a cylindrical self-gravity solver\footnote{https://github.com/wt-liao/gamer\_popIII\_accretion} (see Appendix \ref{S: gamer_cyl} for detail). The Navier-Stokes equation is solved on a uniform cylindrical grid using a Godnov scheme, applying piecewise-linear reconstruction and using an HLLC Riemann solver for the flux prediction. In order to consistently follow the chemistry evolution, chemical abundance is evolved through GRACKLE with 9 species -- electron, HI, HII, HeI, HeII, HeIII, H$^-$, H$_2$, $H_2^+$  \citep{grackle}. For molecular hydrogen formation/dissociation at high density, we use the three body rate from \cite{Glover+08}. The chemical abundances are then used as input to the thermodynamic calculations used in the hydro solver.  Analysis and visualization of the simulations was conducted with \cite{2011ApJS..192....9T}.

\subsection{Initial and Boundary Condition}
We initialize our simulation with a force balanced, axially-symmetric disk with circular rotation. A 0.2 $M_\odot$ central star is sitting at $r=0$, that locates outside the simulation domain and provides a static gravitational potential.
The initial surface density $\Sigma$ and disk midplane temperature $T_0$ are
\begin{eqnarray}
& \displaystyle
\label{eq:Sigma}
\Sigma(r) = \frac{0.01 M_\odot}{\pi (100 {\rm\ AU})^2} \left( \frac{r}{1 {\rm\ AU}} \right)^{-1} \, , \\
& \displaystyle
T_0(r) = 2000 {\rm\ K} \left( \frac{r}{1 {\rm\ AU}} \right)^{-0.15}  \, ,
\end{eqnarray}
which are chosen according to previous cosmological simulation results. 
The initial density field is then seeded with a random perturbation of a magnitude between zero to $10\%$ of the local density, and relaxed for three outer orbits with a constant cooling time $\tau_{\rm cool} = 15 \times \tau_{\rm orbit}$ to allow for the development of turbulence and to avoid artificial fragmentation caused by the initial condition \citep{Deng+17}.  Following this initial relaxation, chemistry is enabled, with
an initial chemical abundance determined by local chemical and ionization equilibrium.  Additionally, the equilibrium is chosen such that the local H$_2$ destruction timescale is equal to the local orbital time.
Following chemical initialization, the system is then relaxed for another outer orbit, during which the cooling rate is increased linearly from zero to the appropriate value.
To ensure the cooling curve is well-sampled, we also require the hydrodynamic time step to be smaller than $5 \%$ of the cooling time.

The simulation domain covers $[0.5 {\rm\ au}, 50{\rm\ au}]$ and $[-50{\rm\ au}, 50{\rm\ au}]$ in the radial and vertical direction respectively, and a full $2 \pi$ in the azimuthal direction. The boundary conditions are periodic in the azimuthal boundary and outflow in the radial boundary. For the vertical boundary condition, the density and pressure are determined using an isothermal and force-balanced boundary. An outflow boundary condition is applied to the velocity fields.
The initial resolution is chosen to be $(N_r, N_\theta, N_z) = (N, 4N, 2N)$, with $N=256$, resulting in a roughly cubical cell at $r = 25 {\rm\ AU}$.
The resolution corresponds to at least $8$ cells per Jeans length for a $\rho=10^{-9} {\rm\ g\ cm^{-3}}$ clump at $10^3 {\rm\ K}$, which is above the Truelove condition, $4$ cells per Jeans length, to avoid artificial fragmentation \citep{Truelove+97}. Past work has suggested that higher resolution is necessary to achieve numerical convergence and to allow for the operation of dynamo processes \citep{Sur+10, Federrath+11, Turk+12}.
The high resolution run would be possible through the adaptive mesh refinement technique, and will be performed in a future study.

\subsection{Optical Depth Approximation}
Molecular hydrogen has two primary cooling channels: one is through ro-vibrational line cooling and the other is through continue cooling triggered by collision, which is commonly known as collision induced emission (CIE). CIE cooling is more efficient at high density, and is typically optically thin when $n \lesssim 10^{16} {\rm\  cm^{-3}}$. In contrast, H$_2$ line cooling becomes optically thick at $n \sim 10^{10} {\rm\ cm^{-3}}$, and thus requires a model for the radiative transfer processes to account for the corresponding cooling. 

To model for the optical depth of H$_2$ line cooling, we apply a fitting function adopted for disk geometry \citep{Liao+19}. We assume a vertically isentropic profile within a disk scale height. The optical depth is:
\begin{eqnarray}
\tau(\eta) =
76 \, f_{\rm H_2} 
\left( \frac{\rho_0}{10^{-10} {\rm\ g\ cm^{-3}}} \right) 
\left( \frac{H}{10 {\rm\ AU}} \right)   
\, \widehat{\tau} \left( T_0, \eta \right) \, ,
\end{eqnarray}
with 
\begin{eqnarray}
\widehat{\tau} \left( T_0, \eta \right) = \widehat{\tau}_0 \, \left( \frac{T_0}{10^3 {\rm\ K}} \right)^{c1}   (1-\eta)^{c2} \, .
\end{eqnarray}
where the coefficients $(\widehat{\tau}, c1, c2) = (1.34, \, -0.79, \, 2.18)$.
In the above equations, we introduce a self-similar disk coordinate $\eta$. At the disk midplane, $\eta \rightarrow 0$; $\eta \rightarrow 1$ corresponds to disk surface. The disk scale height $H = c_s / \Omega$, where $c_s$ is the sound speed at the midplane and $\Omega$ is the angular momentum. $T_0$ is the temperature at midplane. To determine the midplane flow state variables, we utilize the vertically isentropic condition.
In addition, in the disk atmosphere that $z \gg H$, the density barely contributes to the column density. Photon thus travels freely and optical depth becomes much smaller than unity. Accordingly, we set optical depth to a small number when $\vert z \vert > 2 r $.

\begin{figure}[ht!]
\epsscale{1.2}
\plotone{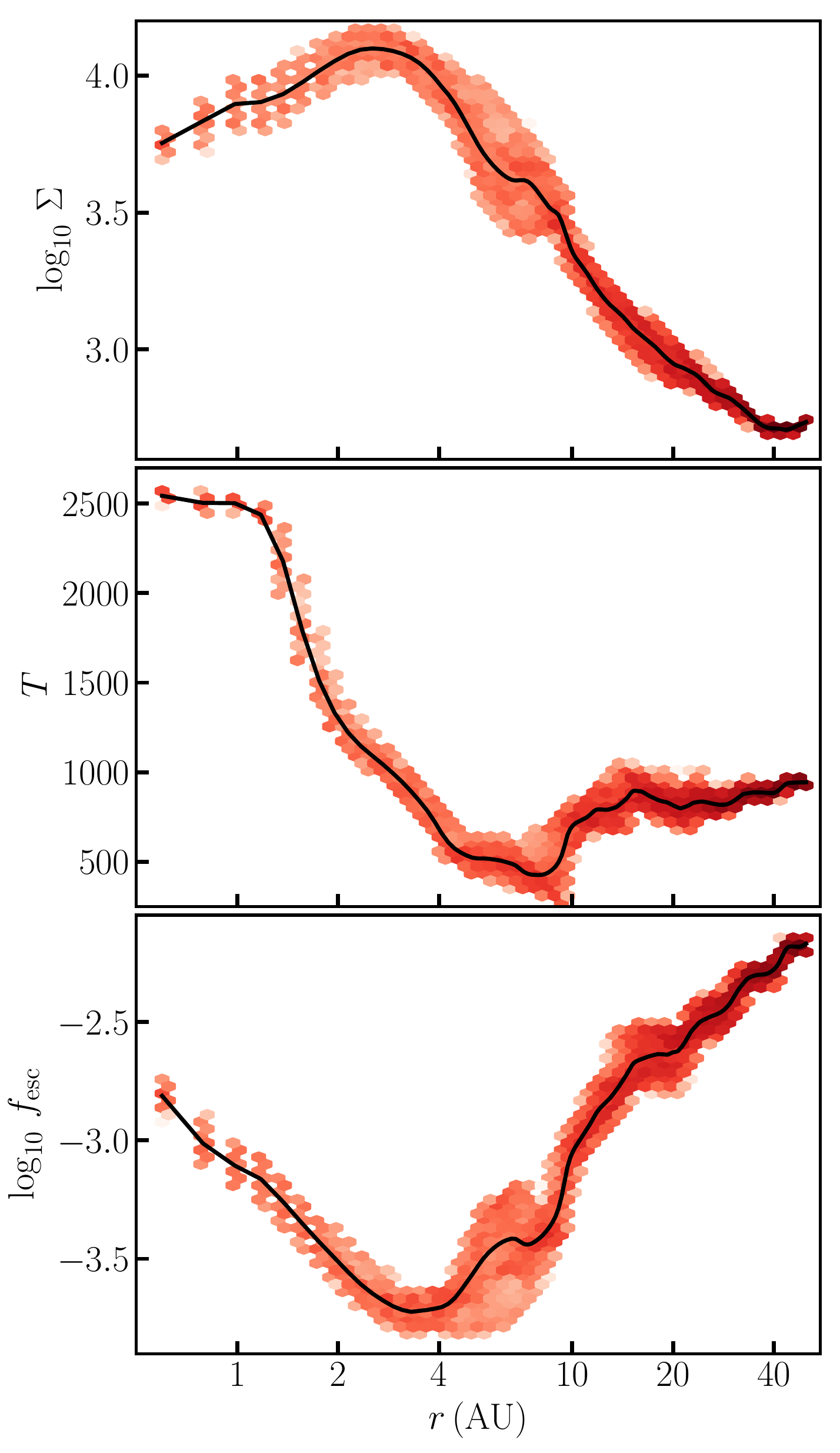}
\caption{Surface density, midplane temperature, and midplane ro-vibrational photon escape fraction at $t=0$. Solid line shows the azimuthally averaged values. The color denotes the probability for a cell to be in the corresponding states, with dark/light indicating high/low probability. \label{fig:f_esc_init}}
\end{figure}

\begin{figure}
\epsscale{1.2}
\plotone{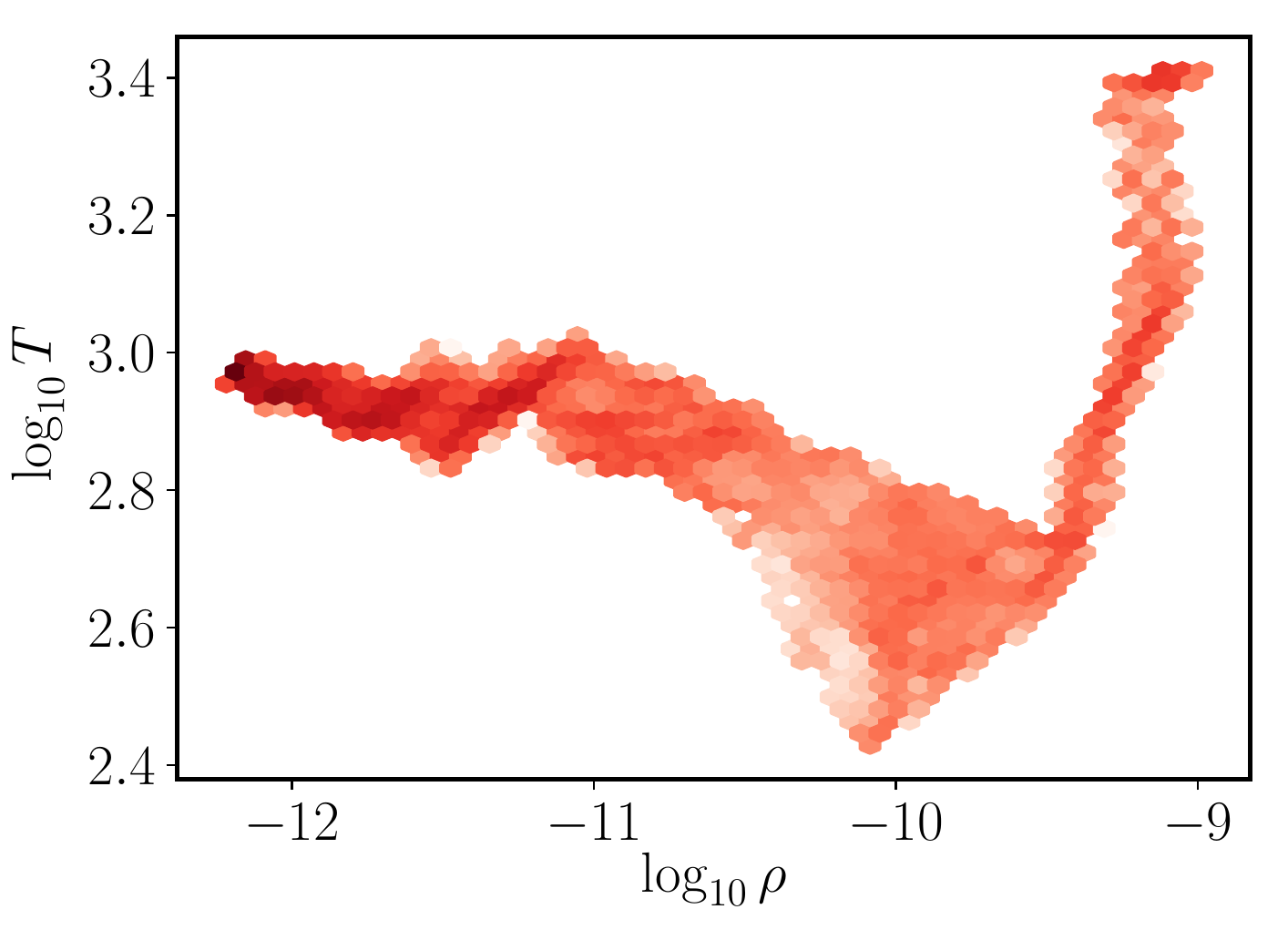}
\caption{Phase diagram (density-temperature) at mid-plane when $t=0$.\label{fig:phase_init}}
\end{figure}

We show surface density, midplane temperature and midplane photon escape fraction as a function of radius in Fig.~ \ref{fig:f_esc_init}, and the midplane density-temperature phase diagram in Fig.~\ref{fig:phase_init}. Note that photon escape fraction is overall smaller than $10^{-2}$, which is caused by large absorption coefficient around $T = 10^3 {\rm\ K}$. We would also like to caution that our optical depth approximation is similar to MH-Ray-M2 method in \cite{Greif+14}, but including the thermal broadening effect. Since the effect of Doppler shift is neglected, photon escape fraction is under-estimated. According to Fig 9 in \cite{Greif+14}, the introduced error would be around a factor of two.


\section{Large Scale Dynamo}
\label{S:dynamo}
The most important ingredient for a large scale dynamo process is the $\alpha_D$ component, with $\alpha_D = \alpha_K + \alpha_M$ \citep{Pouquet+76,Blackman+02}.
In the above expression, $\alpha_K = - \frac{1}{3} \tau_{\rm eddy} \, \overline{ \vec{v'} \cdot ( \vec{\nabla} \times \vec{v'} )  }$ is due to the kinetic heliecity, and $\alpha_M = \frac{1}{3} \rho^{-1} \tau_{\rm eddy} \, \overline{\vec{j} \cdot \vec{b}} $ is originated from the magnetic current helicity, $\vec{j} \cdot \vec{b}$. In the above expressions, $\tau_{\rm eddy}$ is the eddy turn over time, $\vec{v'}$ and $\vec{b}$ represent the turbulent velocity and magnetic fields, and $\vec{j} = \vec{\nabla} \times \vec{b}$.
As a result, at the linear stage, $\alpha_D \approx \alpha_K$ and kinetic helicity is the key ingredient to trigger a large scale dynamo. However, the saturation level depends on the whole $\alpha_D$ expression and thus on the amount of magnetic current helicity retained in the dynamo active region.

In the mean field approximation, a dynamo can be activated from two eigenmodes. 
In the absence of a background mean flow, a pure helical turbulence is able to drive a large scale dynamo; using forced turbulent box simulations, \cite{Brandenburg+01} demonstrates that helical turbulence is able to generate a field beyond the turbulence driving scale, and reach a super-equipartition level. This is known as the $\alpha^2$-dynamo, with a growth rate 
\begin{eqnarray}
\label{eq: gamma_alpha2}
\gamma_{\alpha^2} = \vert \alpha_D^2 k^2 \vert^{1/2} - \left( \eta_t + \eta \right) k^2
\, ,
\end{eqnarray}
where $\eta_t$ is the turbulent diffusion coefficient, $\eta$ is the magnetic diffusivity, and $k$ is the corresponding wave number.

In the presence of a background large scale rotating mean flow, rotation tends to wind up the poloidal field into toroidal components. Subsequently, the feedback from the toroidal field back to the poloidal field is through the $\alpha$-effect. 
This is the well-known $\alpha \Omega$ process. The growth rate of $\alpha \Omega$ process is 
\begin{eqnarray}
\label{eq: gamma_alpha-omega}
\gamma_{\alpha \Omega} =  \frac{1}{2} \vert \alpha_D q_d \Omega k_z \vert^{1/2} - \left( \eta_t + \eta \right) k^2
\, ,
\end{eqnarray}
where $\Omega$ is the disk angular frequency, $q_d = d \ln \Omega / d \ln r$ is the disk shear parameter, and $k_z$ is the wave number along the rotation axis -- z-axis.

\subsection{Dynamo Number}
Large scale dynamo requires a helical turbulent velocity field. 
In an accretion disk, it is commonly believed that turbulence is generated through MRI. However, MRI requires the growth rate to be, at least, comparable to the non-ideal diffusion processes. In a primordial accretion flow, the required minimum field strength is at the order of $10^{-4} {\rm\ G}$, which is several orders of magnitude higher than the field froze in the gas with a primordial origin from, e.g., the Biermann process \citep{Tan+04_dynamo}. 
Numerically, it has also been found that with an initial weak (but finite) and random field, MRI cannot be sustained \citep{Bhat+17}.

However, several other pure hydrodynamic instabilities are known to operate in an accretion disk, e.g., the vertical shear instability (VSI) \citep{Urpin+98,Nelson+13,Richard+16}, and the parametric instability due to the existence of spiral wave \citep{Bae+16a, Riols+17}. 
We suspect that both instabilities are active in the primordial accretion flow, based on the cooling condition, disk geometry and the massive nature of primordial disks.

\begin{figure}
\epsscale{1.1}
\plotone{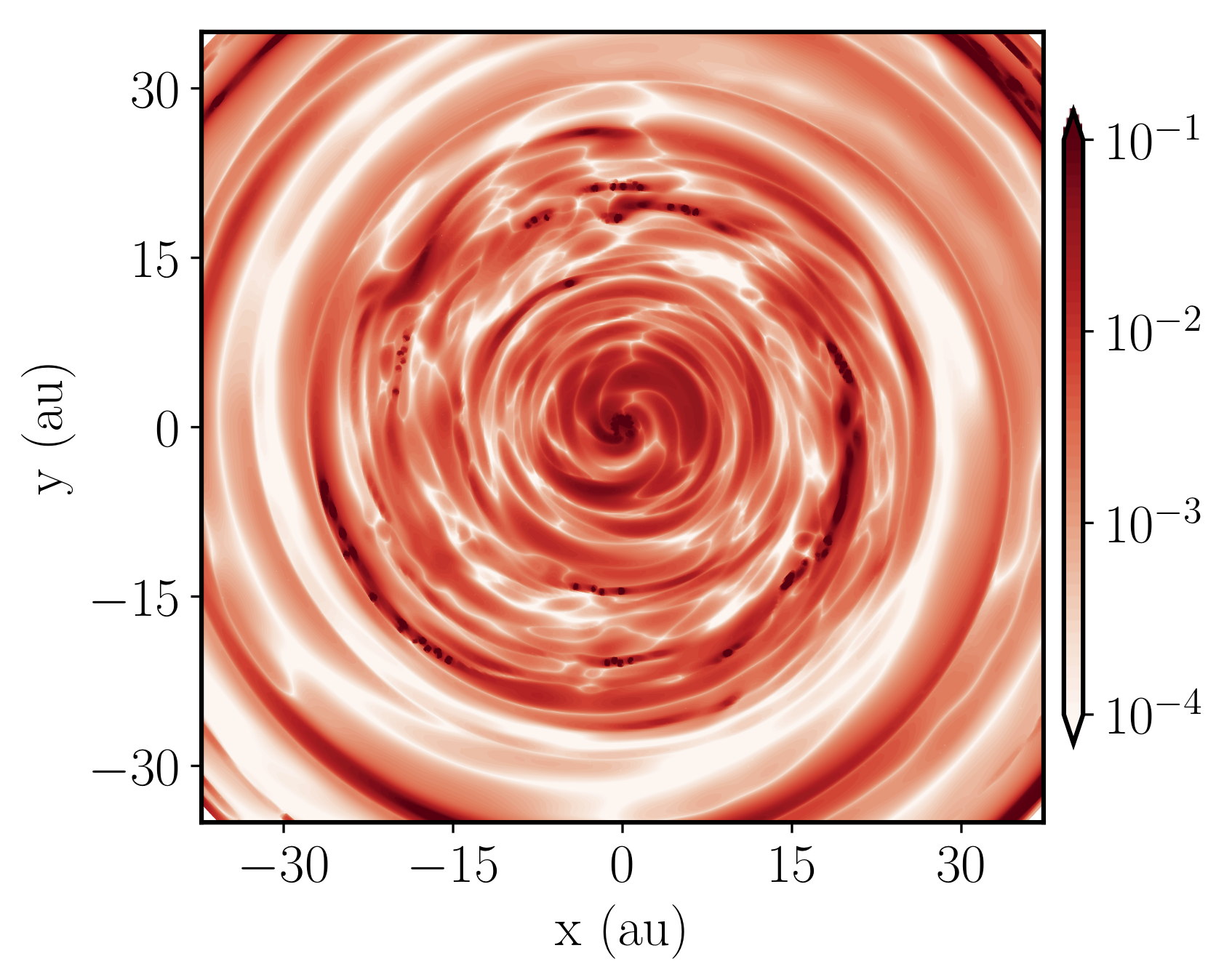}
\caption{Shakura–Sunyaev parameter $\alpha_{ss}$ at the disk midplane after an evolution of one outer orbital time. Primordial accretion flow is turbulent with $\alpha_{ss} \gtrsim 10^{-3}$.
\label{fig:alpha_ss}}
\end{figure}

\begin{figure}
\epsscale{1.1}
\plotone{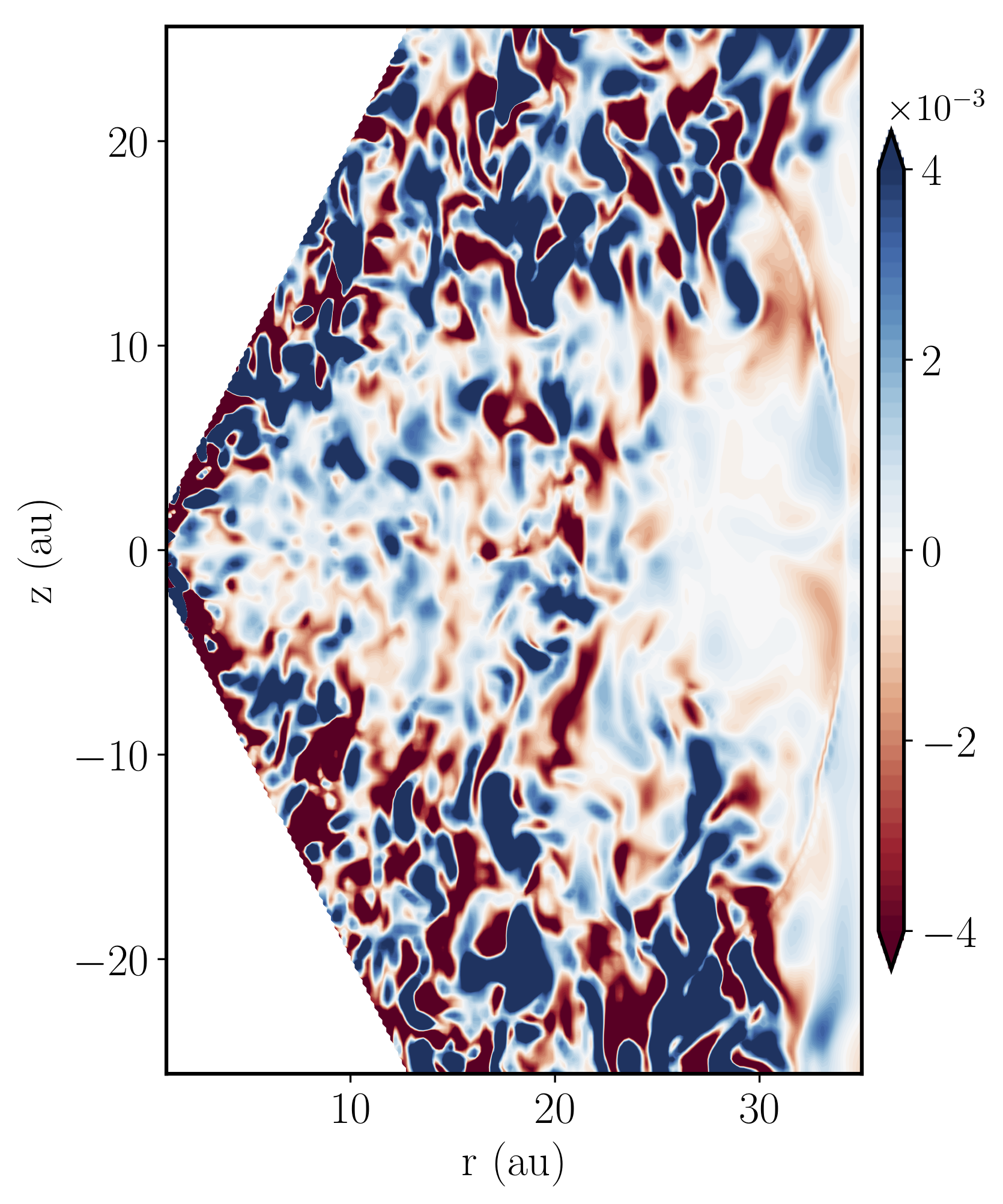}
\caption{Turbulent kinetic helicity $\alpha_K$, in the unit of $1 {\rm\ au}$ and the corresponding orbital time.
Helical turbulence allows large scale dynamo modes to grow.
\label{fig:alpha_mean}}
\end{figure}

Fig \ref{fig:alpha_ss} demonstrates the turbulent field by showing the $\alpha_{ss} = \rho v'_r  v'_\theta / q_d \overline{P}$ from the Shakura-Sunyaev disk model \citep{Shakura+73}, at the disk midplane after an evolution of one outer orbital time. And, $\overline{P}$ is the vertically averaged pressure weighted by density.
We also show the kinetic helical through $\alpha_{K}$ component for a given azimuthal angle in Fig \ref{fig:alpha_mean}.
In deriving the turbulent field, we assume that the background velocity has reached a qusai-steady state and use the azimuthally averaged velocity as the mean flow. 
For simplicity, we use isotropic helicity in $\alpha_K$, and, in Fig \ref{fig:alpha_mean}, normalize $\alpha_K$ with the unit of $1 {\rm\ au}$ and the corresponding orbital time. 
Since the mean flow approximation is more appropriate in the body of a disk, we only show the turbulence related fields at $\vert z \vert \leqq 2 r$.

\begin{figure}
\epsscale{1.1}
\plotone{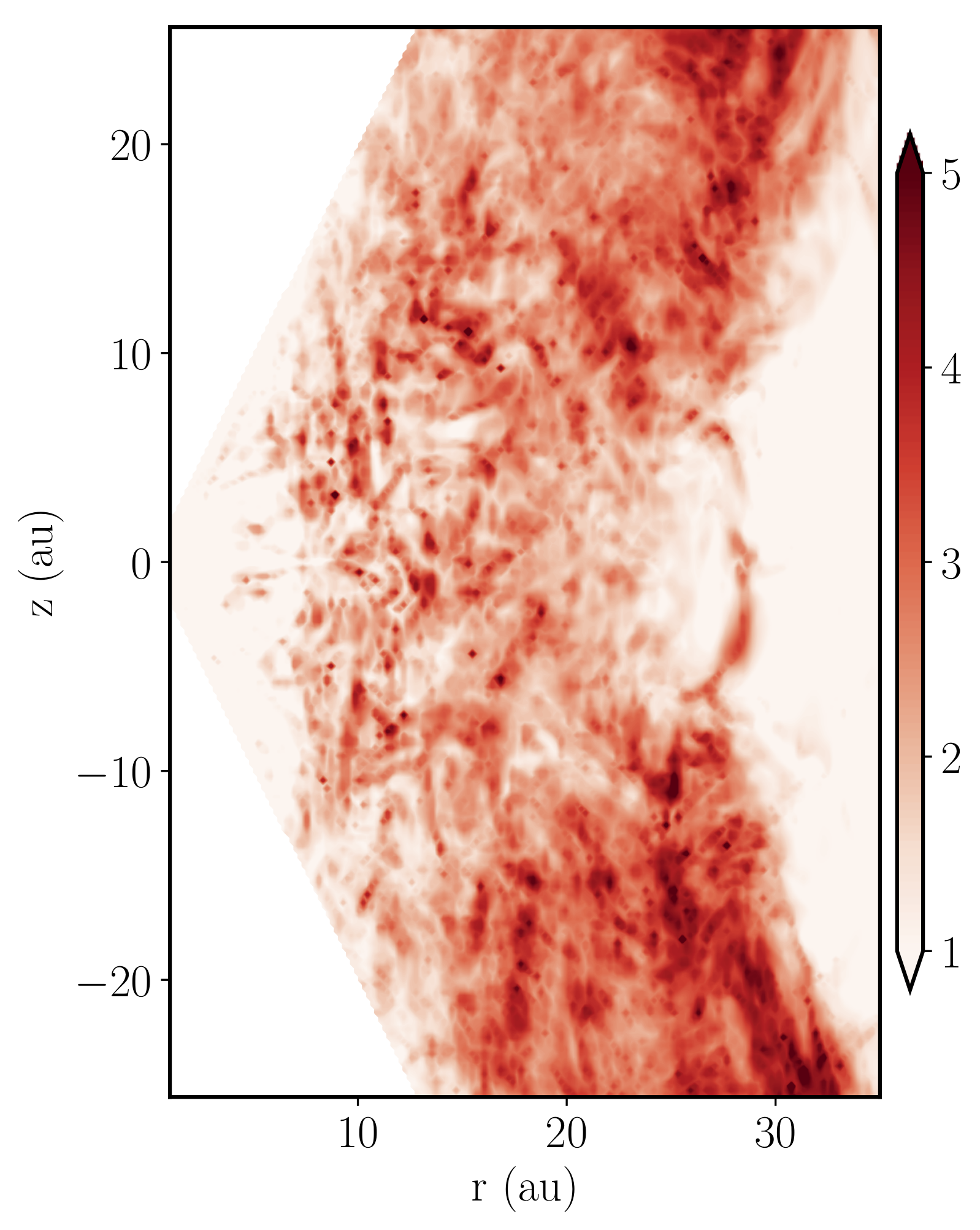}
\caption{Azimuthally averaged $\vert D_{\alpha^2} \vert$, dynamo number of $\alpha^2$-process. For $\vert D_{\alpha^2} \vert > 1$, $\alpha^2$-dynamo is possible at the scales larger than local radius.\label{fig:D_alpha}}
\end{figure}

\begin{figure}
\epsscale{1.1}
\plotone{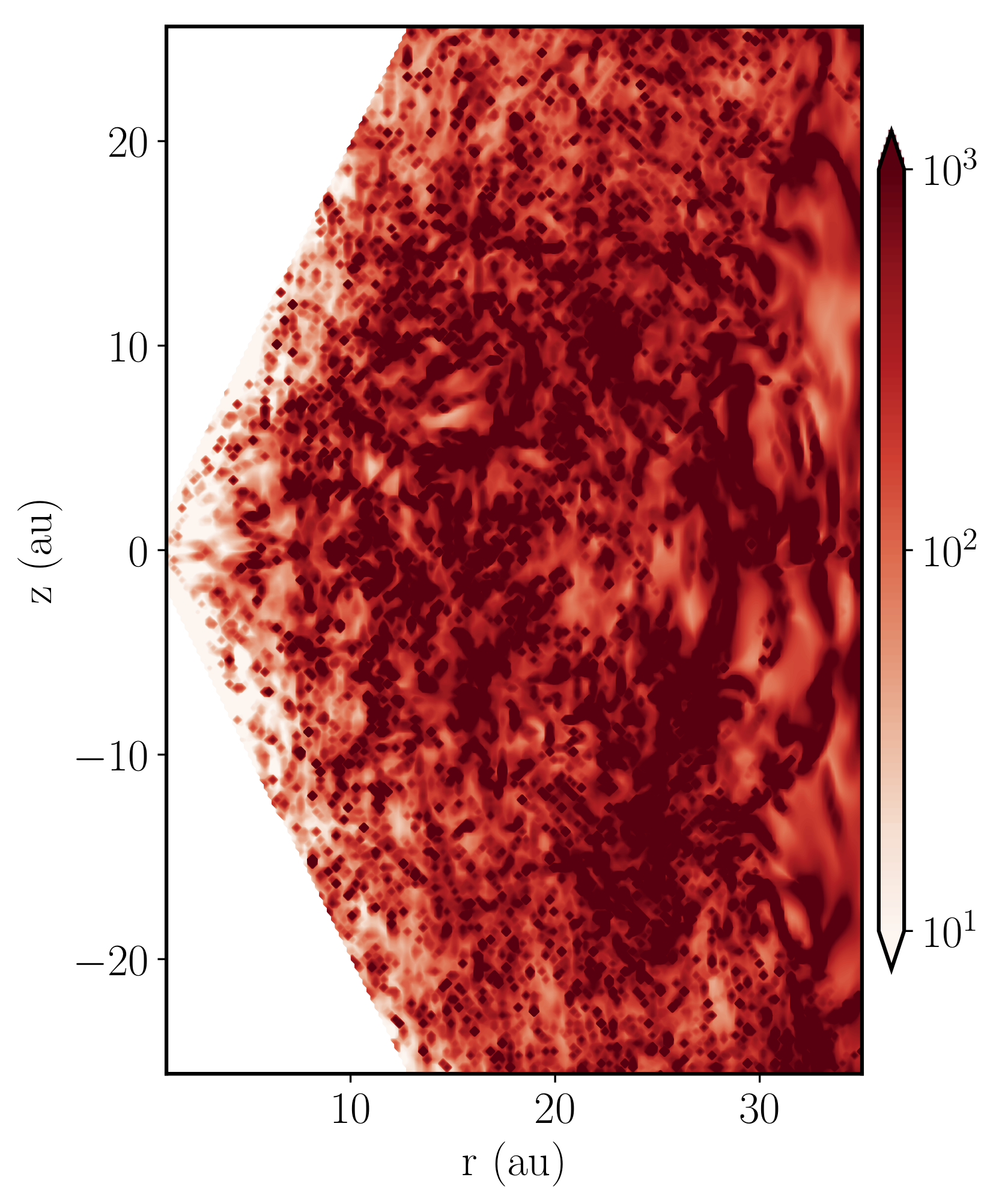}
\caption{Azimuthally averaged $\vert D_{\alpha \Omega} \vert$. Since $\vert D_{\alpha \Omega} \vert \gg 1$, mean field dynamo would be efficient.\label{fig:D_alpha_omega}}
\end{figure}

To activate an $\alpha^2$-dynamo at a given wavenumber $k$, it requires 
$ D_{\alpha^2} = \vert \alpha_D \vert /  \eta_t \, k > 1 $
(cf. Eq \ref{eq: gamma_alpha2}).
%
In Fig \ref{fig:D_alpha}, we show $D_{\alpha^2}(r)$ at $k(r)$, the wavenumber corresponding to the local radius.  For $D_{\alpha^2}(r) > 1$, $\alpha^2$-dynamo is possible at a scale larger than the local radius.
In addition, by requiring the growth rate from Eq.~\ref{eq: gamma_alpha-omega} to be positive, one can derive the expression of dynamo number 
$D_{\alpha \Omega} = 
\alpha \Omega / \eta_t^2 k^3$.
%
With an assumed disk profile, the critical dynamo number is usually found to be $D_{\alpha \Omega, \rm crit} \sim \mathcal{O}(1) - \mathcal{O}(10)$ \citep{Ruzmaikin+88}. We show the azimuthally averaged absolute dynamo number, $ \vert D_{\alpha \Omega} \vert $, in  Fig \ref{fig:D_alpha_omega}. Since $ \vert D_{\alpha \Omega} \vert \gg 10$, it confirms the existence of mean field dynamo in a primordial accretion disk.

\subsection{$Re_M$ and $Pr_M$}
Unlike small scale dynamo that requires a magnetic Reynolds number $Re_M \gtrsim 10^2$, large scale dynamo can be activated with $Re_M \gtrsim 10$ \citep{Brandenburg+01,Iskakov+07,Brandenburg+12}. 
$Re_M$ of a primordial composition has been calculated by, e.g., \cite{Schober+12b, Nakauchi+19}, and is known to be larger than $\mathcal{O}(10^2)$. In the above cited works, the chemical composition and temperature field are derived from the one-zone model. Here, for completeness, we also calculate $Re_M$ using the result from our simulation. 

$Re_M$ is a measure between advection and diffusion dissipation of the magnetic field. It is defined as 
\begin{eqnarray}
Re_M \equiv \frac{c_s H_d }{\eta} \, ,
\end{eqnarray}
where $c_s$ is the local sound speed, $H_d$ is the disk scale height with $H_d \sim r$ for a geometrically thick disk, and $\eta$ is the magnetic diffusivity. 
In principle, the non-ideal effects, including Ohmic dissipation, ambipolar diffusion, and Hall effect, could all contribute to increasing $\eta$. However, ambipolar diffusion and the Hall effect depend on magnetic field strength and are therefore functions of the dynamo processes. 
For simplicity, we consider the contribution from Ohmic dissipation only. We show $Re_M$ in Fig \ref{fig:ReM}, and leave the detail calculation in Appendix \ref{S:ReM}. Indeed, we found $Re_M > 10^2$ in the midplane of the disk, and beyond $\mathcal{O} (10^5)$ at the disk atmosphere. In addition, the fact that $Re_M > 10^4$ in most part of the disk indicates that both small and large scale dynamo are active. Also note that the estimation of $Re_M$ here should be a conservative estimation, since we do not include lithium, a primary electron donor at $T \gtrsim 2000 {\rm\ K}$ \citep{Nakauchi+19}, in our chemistry network.

\begin{figure}
\epsscale{2.2}
\plottwo{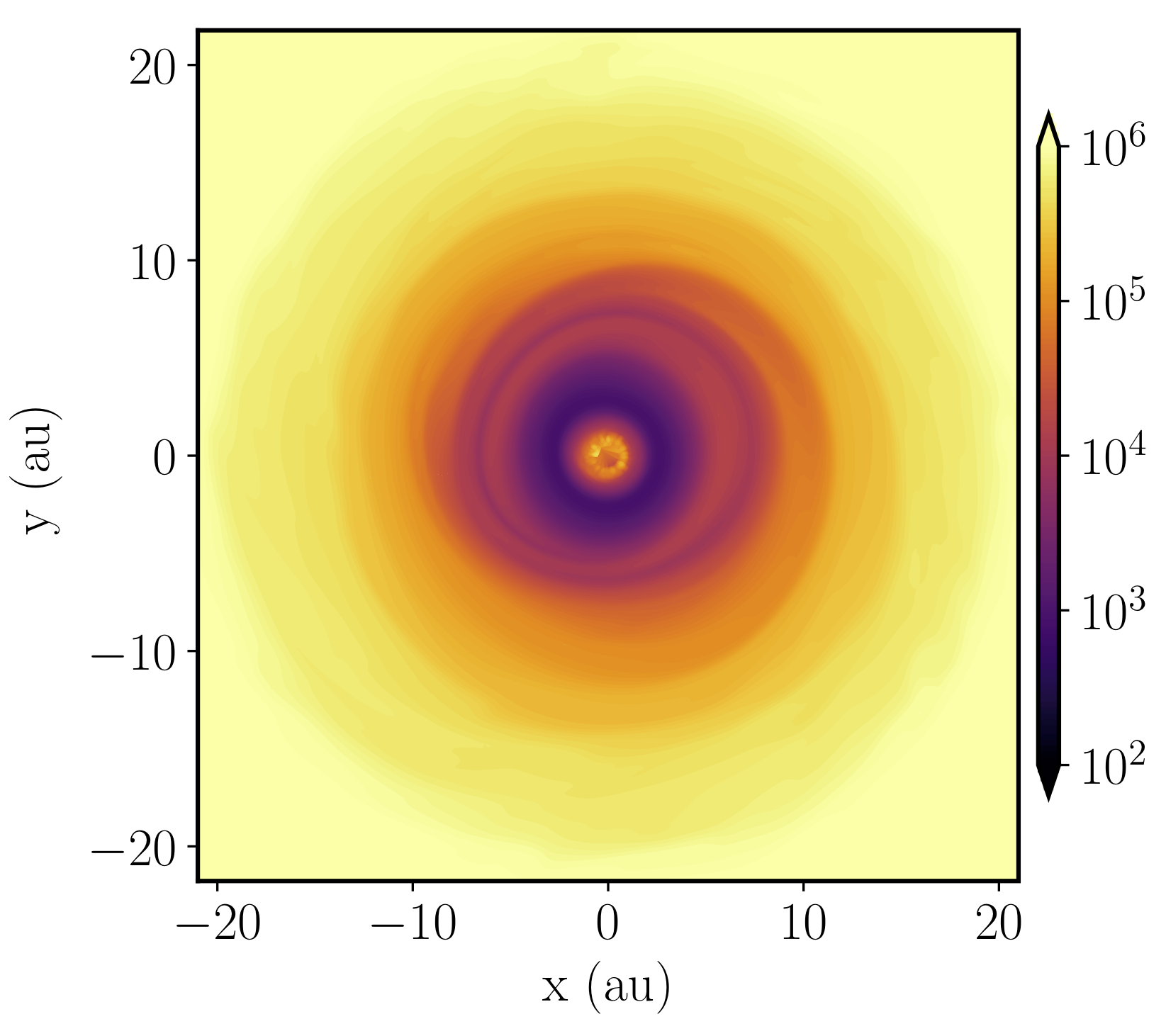}{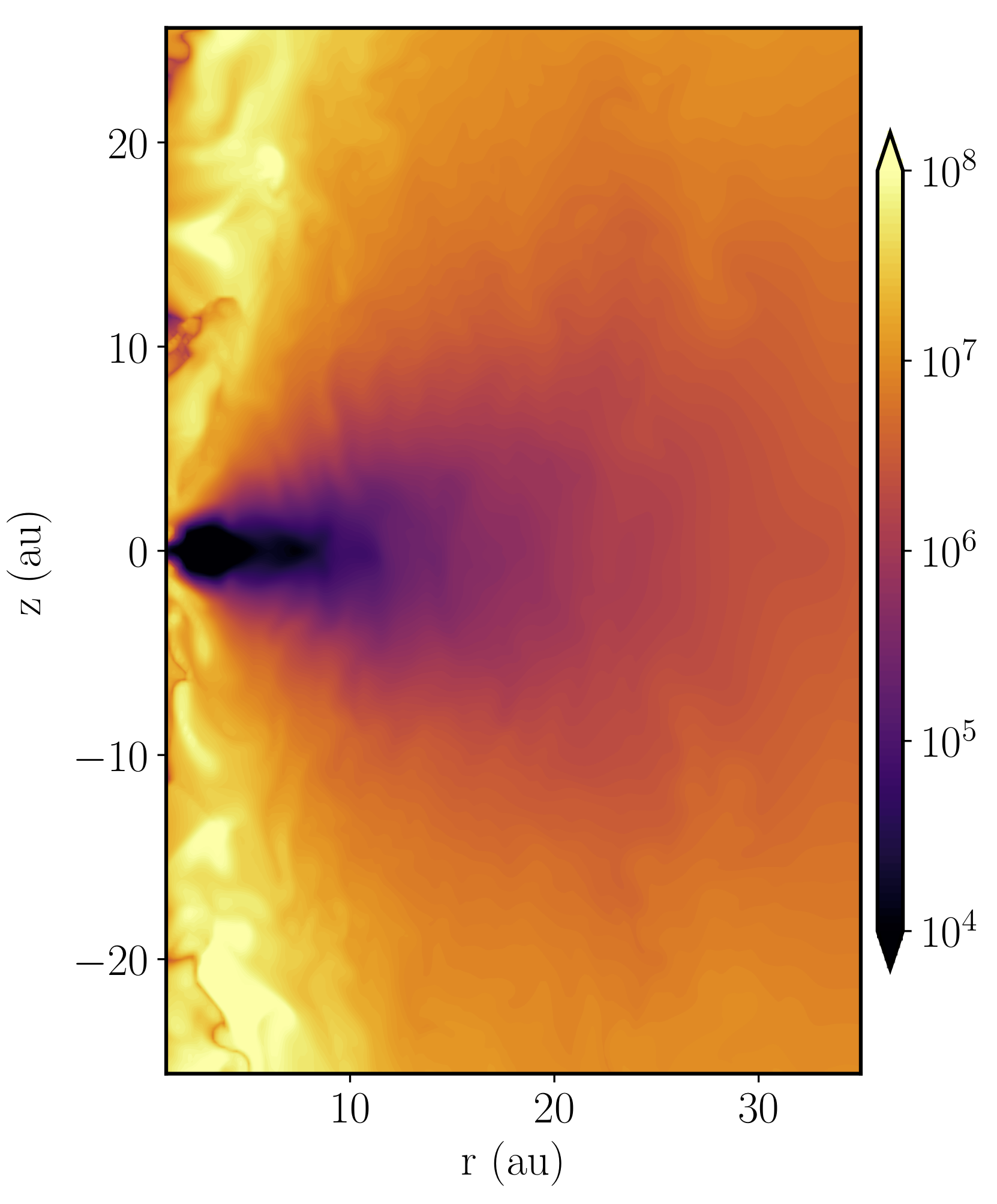}
\caption{Magnetic Reynolds number, $Re_M$ at disk midplane (upper) and at a given azimuthal angle (lower). $Re_M \gg 10$ allows large scale dynamo to activate. \label{fig:ReM}}
\end{figure}
%

\begin{figure}
\epsscale{2.2}
\plottwo{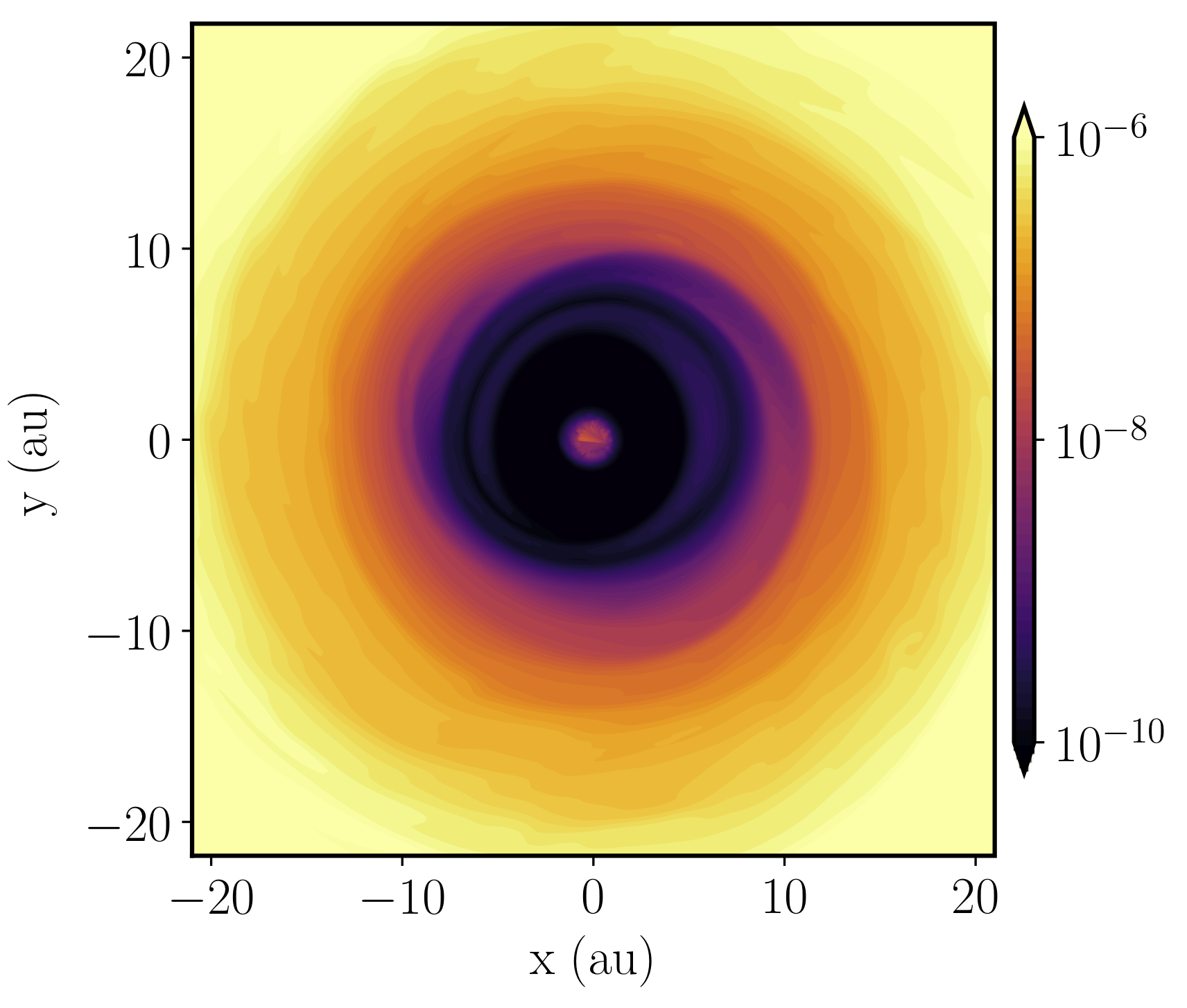}{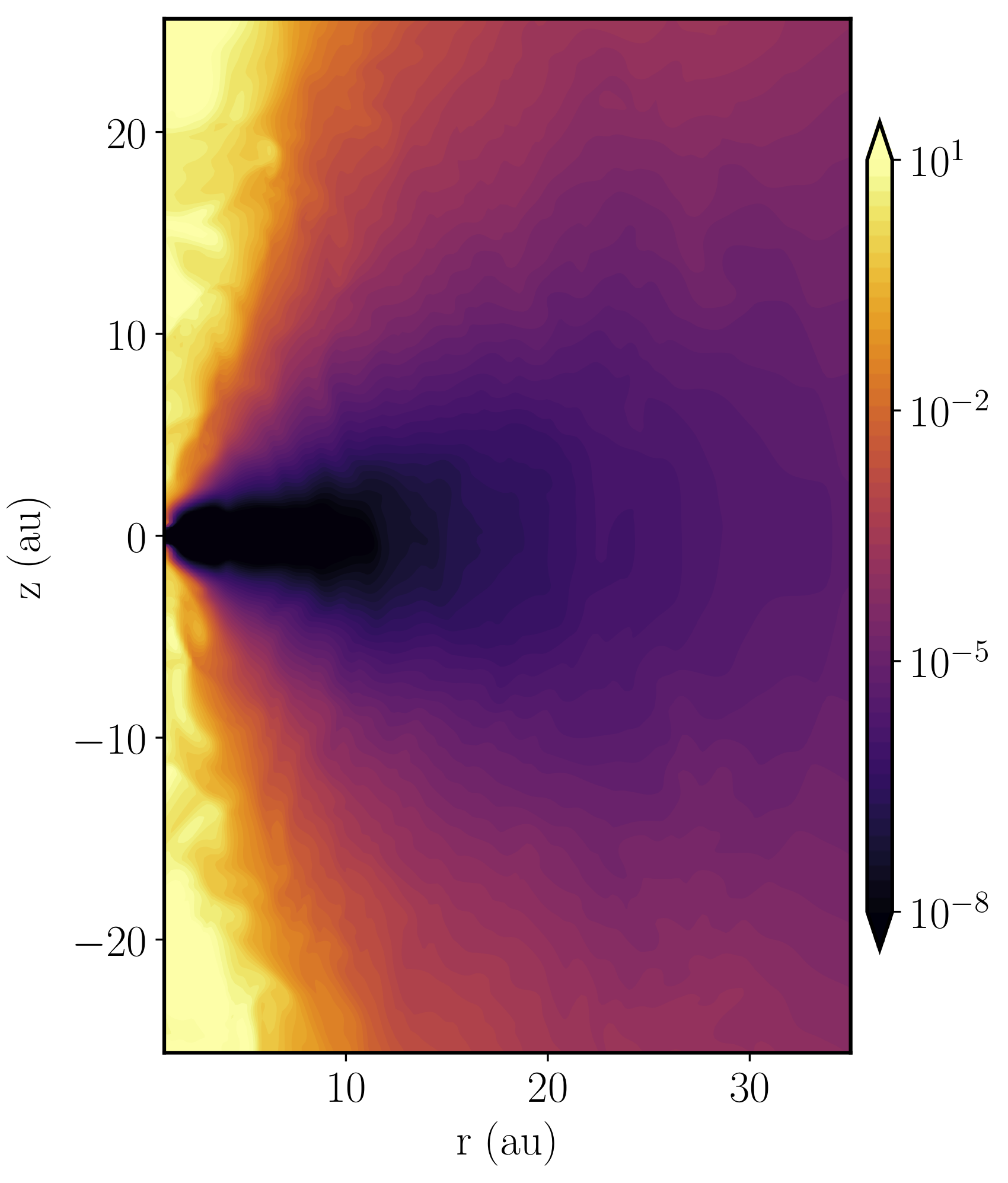}
\caption{Magnetic Prandtl number, $Pr_M$ at disk midplane (upper) and at a given azimuthal angle (lower). $Pr_M \ll 1$ indicates that, through purely small scale dynamo, magnetic field is likely saturated at a level of $\mathcal{O}(10^{-2})$. \label{fig:PrM}}
\end{figure}

Another important parameter is the magnetic Prandtl number, $Pr_M = \nu / \eta$, where $\nu$ is the kinetic viscosity. For $Pr_M \gg 1$, small scale dynamo is easier to be activated and has a higher saturated level, since the stretch and folding is at the scale above the resistive dissipation scale. The critical $Re_M$ also decreases when $Pr_M$ increases \citep{Iskakov+07}.
However, large scale dynamo does not seem to be sensitive to the variation of $Pr_M$ \citep{Brandenburg+09}.
We show $Pr_M$ in Fig \ref{fig:PrM}. The small $Pr_M$ at the disk midplane is because that a high density leads to a small kinetic viscosity and a large magnetic diffusivity.
Since $Pr_M \ll 1$, magnetic field is likely saturate at a level of $\mathcal{O}(10^{-2})$ if the amplification is through small scale dynamo only \citep{Schober+15}.

\subsection{Quenching and Saturation}
The saturation level of large scale dynamo processes depends on the amount of magnetic current helicity retained in the dynamo active region. 
During the large scale dynamo processes, $\alpha_M$ develops together with the amplification of magnetic field. Ultimately, it brings the $\alpha_D \rightarrow 0$, and quenches the large scale dynamo processes. 
A resolution of this is through magnetic helicity flux that brings away small scale $\alpha_M$ and rejuvenates $\alpha_D$  \citep{Blackman+00}. 

\cite{Subramanian+06} provides an expression for the evolution of magnetic helicity density ($h_M$) due to the helicity flux ($\mathcal{\vec{F}}_h$):
\begin{eqnarray}
\label{eq:helicity_flux1}
\frac{\partial h_M}{\partial t} + 
\vec{\nabla} \cdot \mathcal{\vec{F}}_h = 
- 2 \mathcal{\vec{E}} \cdot \vec{B}
- 2 \eta \overline{ \vec{j} \cdot \vec{b} }
\, .
\end{eqnarray}
In the above equation, $\mathcal{\vec{E}} = \overline{\vec{v} \times \vec{b} } = \alpha \vec{B} - \eta_t \vec{J} $ is the turbulent electromotive force (EMF), and the last equality is valid under the framework of a mean field dynamo and neglects the higher order terms.
Following \cite{Shukurov+06}, Eq \ref{eq:helicity_flux1} can be re-written as 
\begin{eqnarray}
\label{eq:helicity_flux2}
\frac{\partial \alpha_M}{\partial t} 
\approx
- \frac{2 \eta_t}{l_0^2} 
\left( 
\alpha_D \hat{B}^2
+
\alpha_M
\left(
\hat{k}_f^2 + \hat{\eta}_t^{-1}
\right)
\right)
-
\vec{\nabla} \cdot 
\left(
\alpha_M \vec{U}
\right)
\, ,
\end{eqnarray}
where $\vec{U}$ represents the large scale flow that brings $\alpha_M$  away from the dynamo active zone, $l_0$ is the integral scale of the turbulence, $\hat{B}^2 = B^2 / B_{eq}^2$, $B_{eq} = \rho \overline{v'^2}$, $\hat{k_f} = k_f / k_0$, $k_f$ is the wavenumber corresponding to the driving force (i.e., at the scale of the disk scale height), and $\hat{\eta}_t = \eta_t / \eta$.

In a primordial accretion disk, the large scale flow could be from the accretion, which brings some amount of the $\alpha_M$ away from the bulk of the disk -- dynamo active zone -- to the protostar where strong differential rotation would create a different dynamo condition. 
At the linear stage, the dynamo is governed by the kinetic helicity, and is gradually quenched due to the development of current helicity.
At steady state, Eq \ref{eq:helicity_flux2} becomes
\begin{eqnarray}
\label{eq:helicity_flux3}
\hat{B}_{\rm sat}^2 \sim 
\vert
\frac{\hat{\alpha}_K}{D_{\alpha \Omega}/Re_{M,t}^2}
\vert
\left( 
\hat{k}_f^2  
+ 
\hat{k}_f^2 \hat{U} Re_{M,t}
+ 
\hat{\eta}_t^{-1}
\right)
\, ,
\end{eqnarray}
where $\hat{\alpha}_i = \alpha_i / c_s $, $\hat{U} = U /c_s$, and $Re_{M, t} = c_s H_d / \eta_t = Re_M / \hat{\eta_t}$. 
We also assume that, at saturation, $\vert \alpha_D \vert \ll \vert \alpha_K \vert$ and $\vert \alpha_M \vert \sim \vert \alpha_K \vert$.
Using the mean value from the simulation result at the midplane with $10 < r {\rm\ (au)} < 20 $, we have $\vert \hat{\alpha}_K \vert \rightarrow 0.01$, $\hat{\eta}_t \rightarrow 250$, $Re_M \rightarrow 4 \times 10^5$, so $Re_{M, t} \rightarrow 1.6 \times 10^3$.
Typically, $\hat{k}_f \sim \mathcal{O}(10^{-1})$, and we choose $( \hat{k}_f^2 + \hat{k}_f^2 \hat{U} Re_{M,t} + \hat{\eta}_t^{-1} ) \rightarrow \hat{\eta}_t^{-1}$ for a conservative estimation of $\hat{B}_{\rm sat}$. The above approximation also assumes a slow accretion with $\hat{U} \ll 1$. 
With $D_{\alpha \Omega} \sim D_{\rm crit} = 10$ at the saturation, we have $\hat{B}_{\rm sat} \sim 3 > 1$. 
The large scale dynamo is able to amplify the magnetic field to an equipartition level in a primordial accretion flow. 
And, the timescale of a large scale dynamo is the dissipation time, $\tau_{\alpha \Omega} \sim H_d^2 / \eta_t \sim Re_{M,t} \Omega^{-1}$.

\section{Discussion \& Conclusion}
\label{S: conclusion}
Based on the turbulent velocity field from pure hydrodynamic simulaion results, we investigate the likelihood of observing large scale dynamo processes in a primoridal accretion flow. We demonstrate that, as the accretion flow is turbulent with $\alpha_{ss} \gtrsim 10^{-3}$, and has $D_{\alpha \Omega} \gg 10$, a fast, large scale dynamo can result during the linear stage of evolution. 

In the presence of a large-scale dynamo, the saturation level depends on the quantity of current helicity retained in the dynamo-active regions. 
Under a slow accretion, with $\hat{U} \ll 1$ that may be able to prevent the dynamo from being completely quenched, we estimate that the saturation level would have $B/B_{\rm eq} \sim 3$. This estimation provides a point of view at the saturation when $\vert \hat{\alpha}_D \vert \ll 1$ but $\hat{\alpha}_D \neq 0$.
We note that for an accretion rate of $\dot{M} \sim 10^{-3} M_\odot {\rm yr^{-1}}$, $\hat{U} \sim \mathcal{O}(10^{-1})$; as a result, the helicity flux term in Eq.~\ref{eq:helicity_flux1} may quickly re-distribute $\alpha_M$. A self-consistent calculation that includes the dynamo equations is necessary to determine the evolution of the magnetic field, although our calculation is able to place constraints.

Our analysis also relies on the assumption that the turbulent statistics remain the same when magnetic fields are present in the accretion disk. 
While this may be true during the kinetic stage of dynamo processes, as the magnetic field is (by definition) too weak to be dynamically important, if a coherent field with $ B \gtrsim 10^{-4} {\rm\ G}$ is attained, the MRI would also be active in the accretion disk \citep{Tan+04_dynamo}. This could then become the primary source of turbulence, and furthermore, a strong magnetic field would suppress both MRI and VSI \citep{Urpin+98}. If there is no other active instability in the strong magnetic field limit, the saturation level is unlikely to exceed equipartition.

If the field saturation level reaches equipartition, it may be able to efficiently brake the rotation of PopIII stars \citep{Hirano+18}. It affects the efficiency of rotational mixing inside the star, and thus the elements that would be produced during the stellar evolution. 
In addition, the presence of a strong, coherent field would favor asymmetric supernovae \citep{Ezzeddine+19}. 
If the parent dark matter halo ultimately serves as the host of a supermassive black hole, the strong coherent field would also be able to produce sychrontron radiation that could explain for the detected excess radio background \citep{Feng+18}, and thus the global 21-centimeter absorption depth at $78 {\rm\ MHz}$ \citep{Bowman+18}.
Although the dynamo problem is numerically difficult to solve since the most efficient dynamo mode locates at the dissipation scale that is beyond the resolution limit, future numerical studies that include magnetic field would still be beneficial to understand the linear phase, the impact from field topology, the helicity flux and the turbulence statistics.

\acknowledgments
W.-T. L. and M.J.T. are supported in part by the Gordon and Betty Moore Foundation’s Data-Driven Discovery Initiative through Grant GBMF4561 to Matthew Turk. W.-T. L. acknowledges government scholarship to study aboard from the ministry of education of Taiwan.
H.S. acknowledges the funding support from the Jade Mountain Young Scholar Award no. NTU-108V0201, MOST of Taiwan under the grant no. 108-2112-M-002-023-MY3, and the NTU Core Consortium project under the grant no. NTU-CC-108L893401.
This research is part of the Blue Waters sustained-petascale computing project, which is supported by the National Science Foundation (awards OCI-0725070 and ACI-1238993) and the state of Illinois. Blue Waters is a joint effort of the University of Illinois at Urbana-Champaign and its National Center for Supercomputing Applications.

\appendix
\section{Cylindrical Solver in GAMER}
\label{S: gamer_cyl}
In this section, we demonstrate the accuracy of the cylindrical fluid and self-gravity solvers, and leave the detail implementations and scaling tests to a future discussion.
The cylindrical fluid solver aims to conserve the angular momentum, which is similar to the implementation in, e.g., Athena \citep{athena_cyl} and PLUTO \citep{pluto}.
All the test problems presented here use piecewise-linear reconstruction and HLLC Riemann solver for the flux prediction -- the same numerical schemes used in the production run.

\begin{figure}[ht!]
\epsscale{0.55}
\plotone{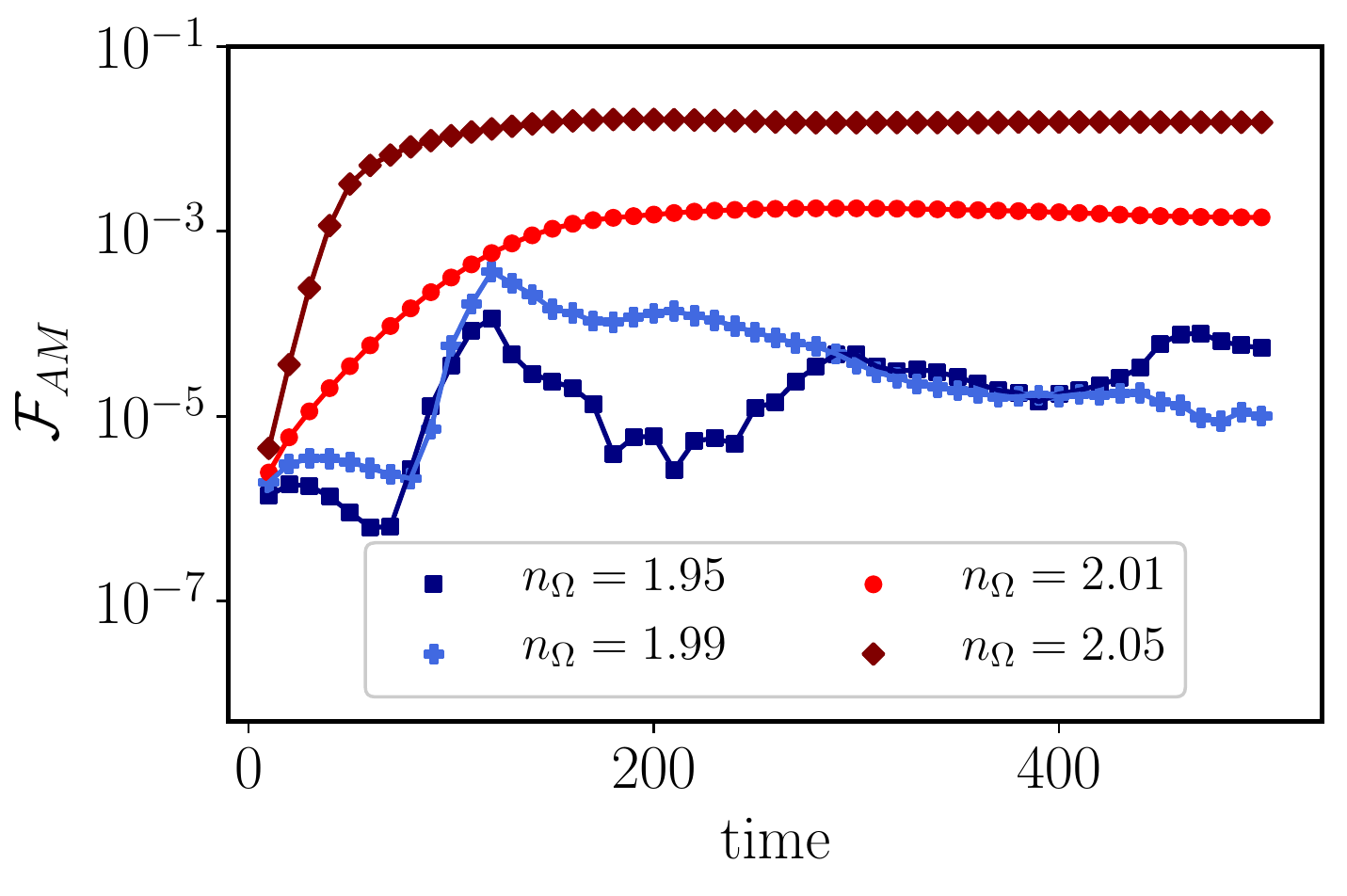}
\caption{Angular momentum flux, $\mathcal{F}_{AM}$, for a rotating flow with different equilibrium velocity profiles.
\label{fig:rayleigh}}
\end{figure}

The first test problem demonstrates the capability of angular momentum conservation by examining the Rayleigh's criterion.
Rayleigh's criterion requires the angular momentum to increase outward, i.e., $\partial \left( r^2 \Omega(r) \right)/\partial r > 0$.
For $\Omega(r) \propto r^{- n_\Omega}$, $n_\Omega < 2$ is the stability condition. For $n_\Omega > 2$, a perturbation of $\Omega$ would trigger angular momentum exchange between different annuli.
In Fig \ref{fig:rayleigh}, we show the angular momentum flux $\mathcal{F}_{AM} = \int \rho r v_r \delta v_\theta \, dV / \int R P \, dV$, where $\delta v_\theta = v_\theta - r \Omega_0(r)$ and $\Omega_0(r)$ is the background equilibrium angular velocity \citep{athena_cyl}.
In the test problem, $\rho = 100$, $P = 1$, and $v_\theta(r=1)=1$, and we perturb the angular velocity $\Omega(r)$. The simulation box has a size of $[1,5] \times [0, 2 \pi] \times [0, 1]$, and has a resolution of $(64, 128, 16)$ in $(r, \theta, z)$ directions.
For $n_\Omega > 2$, the unstable velocity profile initiates angular momentum exchange, whereas, for $n_\Omega < 2$, the angular momentum flux is fluctuated around the perturbation level.

\begin{figure}[ht!]
\epsscale{1.2}
\plotone{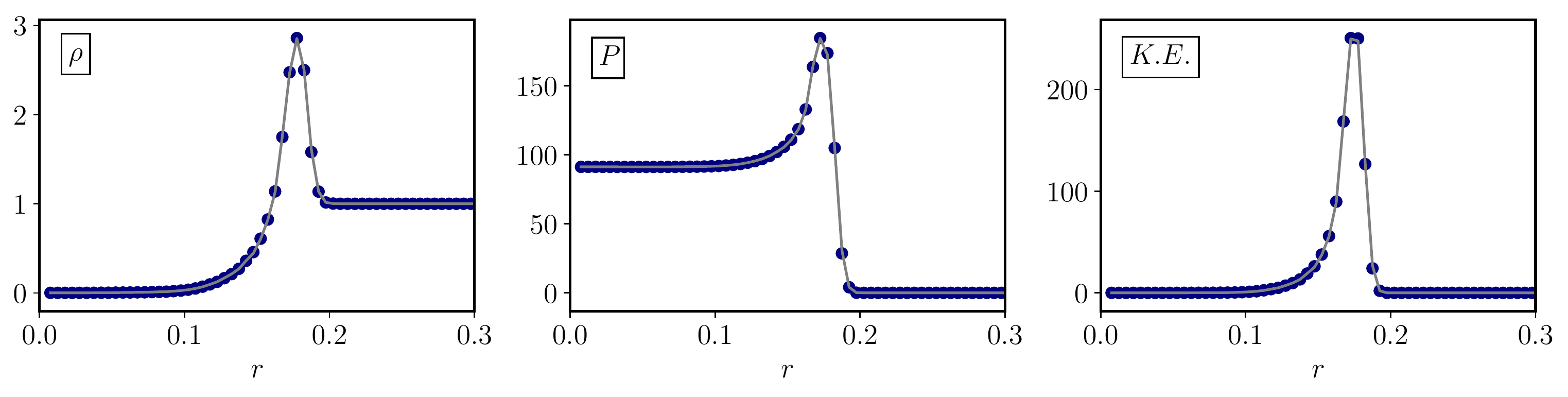}
\caption{The density ($\rho$), pressure ($P$) and kinetic energy ($K.E.$) of the blast wave test problem solved in both Cartesian coordinate (gray solid line) and cylindrical coordinate (blue circles).
\label{fig:blast}}
\end{figure}

As another test problem, in Fig \ref{fig:blast}, we show the solutions of the blast wave problem solved in both Cartesian (shown in gray solid line) and cylindrical coordinate (shown in blue circles).
The values shown in Fig \ref{fig:blast} are averaged over the same radial distance from the blast center.  
In the test problem, the total energy and the initial radius of the blast wave are: $E_{\rm blast} = 10$ and $r_{\rm blast} = 0.1$. The background density is set to be unity and background energy density is $2.5 \times 10^{-5}$. 
The simulation box has a size of $[0.5,1.5] \times [0, 1] \times [0, 1]$ in the $(r, \theta, z)$ or $(x, y,z)$ directions. The resolution of both run are $128^3$.

\begin{figure}[ht!]
\epsscale{1}
\plottwo{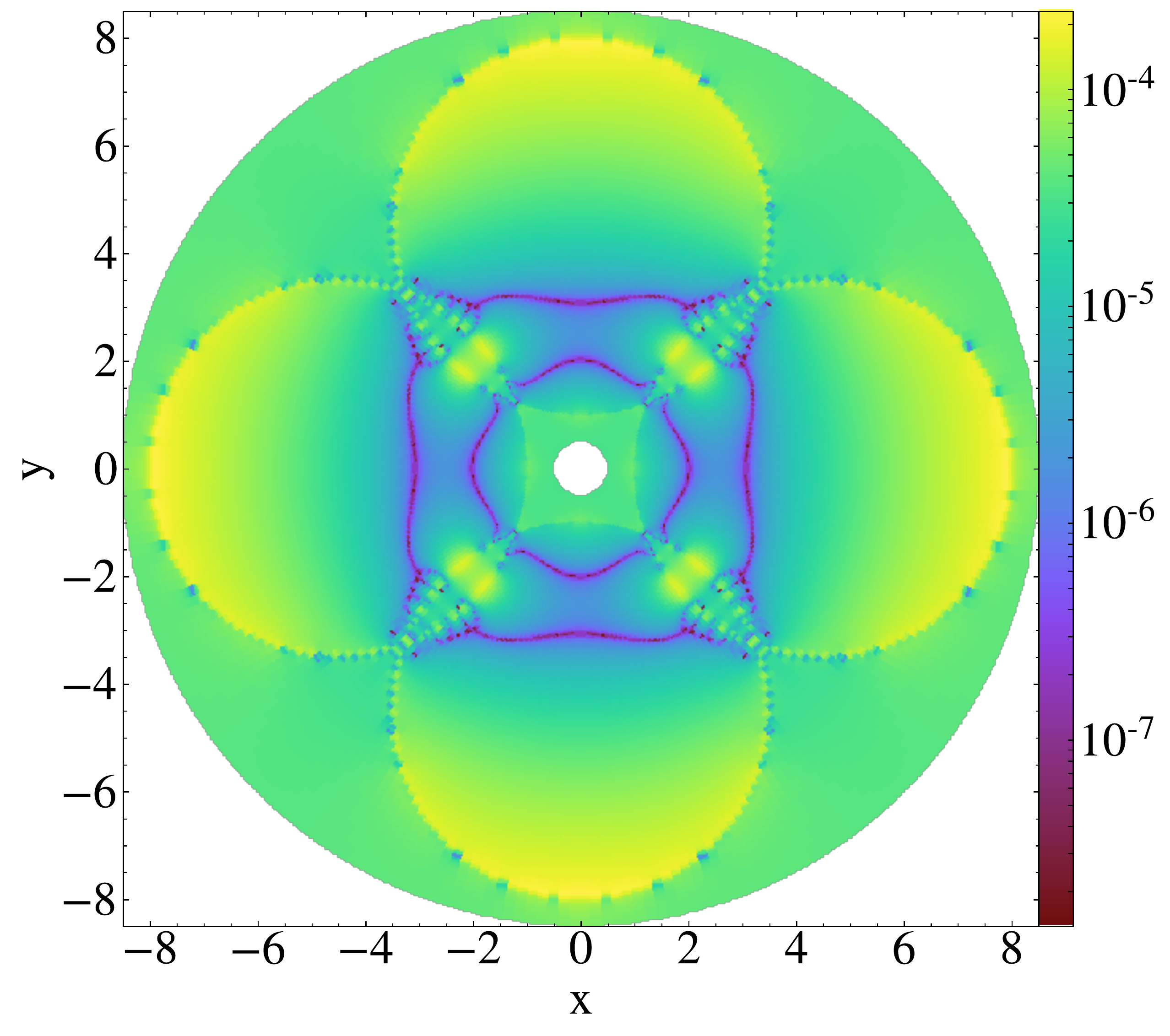}{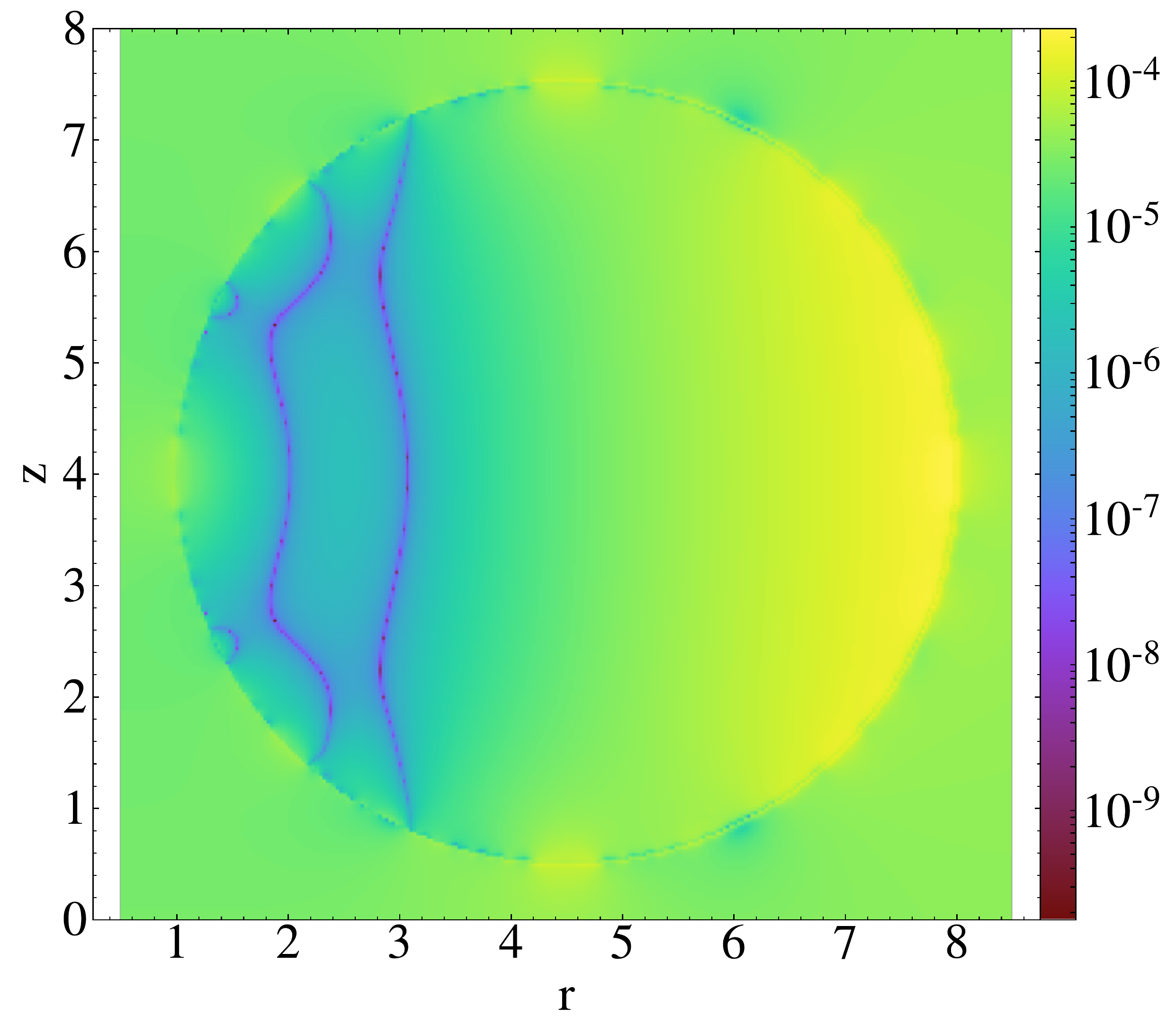}
\caption{Relative error (lower is better) of the gravitational potential in the four sphere test problem with a resolution of $N=256^3$. The left panel shows the relative error map at $z=4$ plane; The right panel shows the map at $\theta =0$ plane. Both maps cut through the center of the sphere(s).
\label{fig:4sphere}}
\end{figure}

For the self-gravity solver in cylindrical coordinate, we solve the integral form of the Poisson equation, and use fast Fourier transform in the $\theta$ and $z$ directions. 
As a test problem, we solve the gravitational potential of four uniform spheres. Each of them has a density $\rho_0=10$ and radius $R_0=3.5$. 
The centers of the spheres are located at $(4.5, \, 0.0, \, 4.0)$, $(4.5, \, 0.5 \pi, \, 4.0)$, $(4.5, \, 1.0 \pi, \, 4.0)$, and $(4.5, \, 1.5 \pi, \, 4.0)$. 
The simulation domain is $[0.5, \, 8.5] \times [0, \, 2 \pi] \times [0, \, 8]$. 
The gravitational potential of each sphere is 
\begin{eqnarray}
\label{eq: sphere}
\Phi_{\rm sphere}(R) = 
\left\lbrace
\begin{array}{ll}
\displaystyle
- \frac{2 \pi G \rho_0}{3} \left( 3 R_0^2 - R^2 \right) & {\rm\ ,if\ } R \leq R_0 \\
\displaystyle
- \frac{4 \pi G \rho_0}{3} \frac{R_0^3}{R} & {\rm\ ,if\ } R > R_0 
\end{array} \right . 
\, ,
\end{eqnarray}
and the final resulting potential is the superposition of all spheres.
In Fig \ref{fig:4sphere}, we show the relative error of the four sphere problem.

\section{Magnetic Reynolds Number}
\label{S:ReM}
Magnetic Reynolds number, to re-iterate,
\begin{eqnarray}
Re_M \equiv \frac{c_s H_d}{\eta} \, ,
\end{eqnarray}
where $c_s$ is the sound speed, $H_d$ is the disk scale height, $\eta \equiv c^2 / 4 \pi \sigma$ is the magnetic diffusivity, and $\sigma$ is the conductivity. By definition \citep{Nicholson+83}, 
\begin{eqnarray}
\sigma \equiv \frac{n_e c^2}{m_e \nu_{\rm coll}} \, ,
\end{eqnarray}
where we have assumed electron is the primary current carrier because $m_e / m_p \ll 1$, and $n_e$ is the electron number density, $m_e$ ($m_p$) is the electron (proton) mass, $\nu_{\rm coll}$ is the electron collision frequency. 
Electron colliding with either another electron/ion and with a neutral particle would both impede its mobility and ultimately contribute to the conductivity. Therefore, 
\begin{eqnarray}
\nu_{\rm coll} = \nu_{\rm coll, \, ei} + \nu_{\rm coll, \, en} 
\, ,
\end{eqnarray}
where $\nu_{\rm coll, \, ei}$ and $\nu_{\rm coll, \, en}$ represent the collision frequency between electron-ion and electron-neutral respectively. 
We also note that the collision frequency between electron-electron and electron-ion is at the same order, so it is sufficient to discuss one of them.

For collision between electron and neutral, 
\begin{eqnarray}
\nu_{\rm coll, \, en} = n_n 
\left\langle \sigma v \right\rangle 
\, ,
\end{eqnarray}
where $\left\langle \sigma v \right\rangle$ is the coefficient of momentum transfer rate, and is taken from \cite{Pinto+08}. 
For collision between electron and ion, we have \citep{Nicholson+83}
\begin{eqnarray}
\nu_{\rm coll, \, ei} 
= 
\frac{8 \pi n_i e^4}{m_e^{1/2} \left( k_B T\right)^{3/2}} \ln \Lambda
\, ,
\end{eqnarray}
where $n_i$ is ion number density, $e$ is the electron charge, $k_B$ is the Boltzmann constant, $\Lambda = n_e \lambda_D^3$ is the plsmaa parameter, and $\lambda_D = \sqrt{k_B T / 8 \pi n_e e^2}$ is the Debye length. 
We have also assumed that the integral of small angle scattering effect between charged particles is more significant than the large angle scattering, which is true when $\Lambda \gg 1$.

\begin{figure}[ht!]
\epsscale{1}
\plottwo{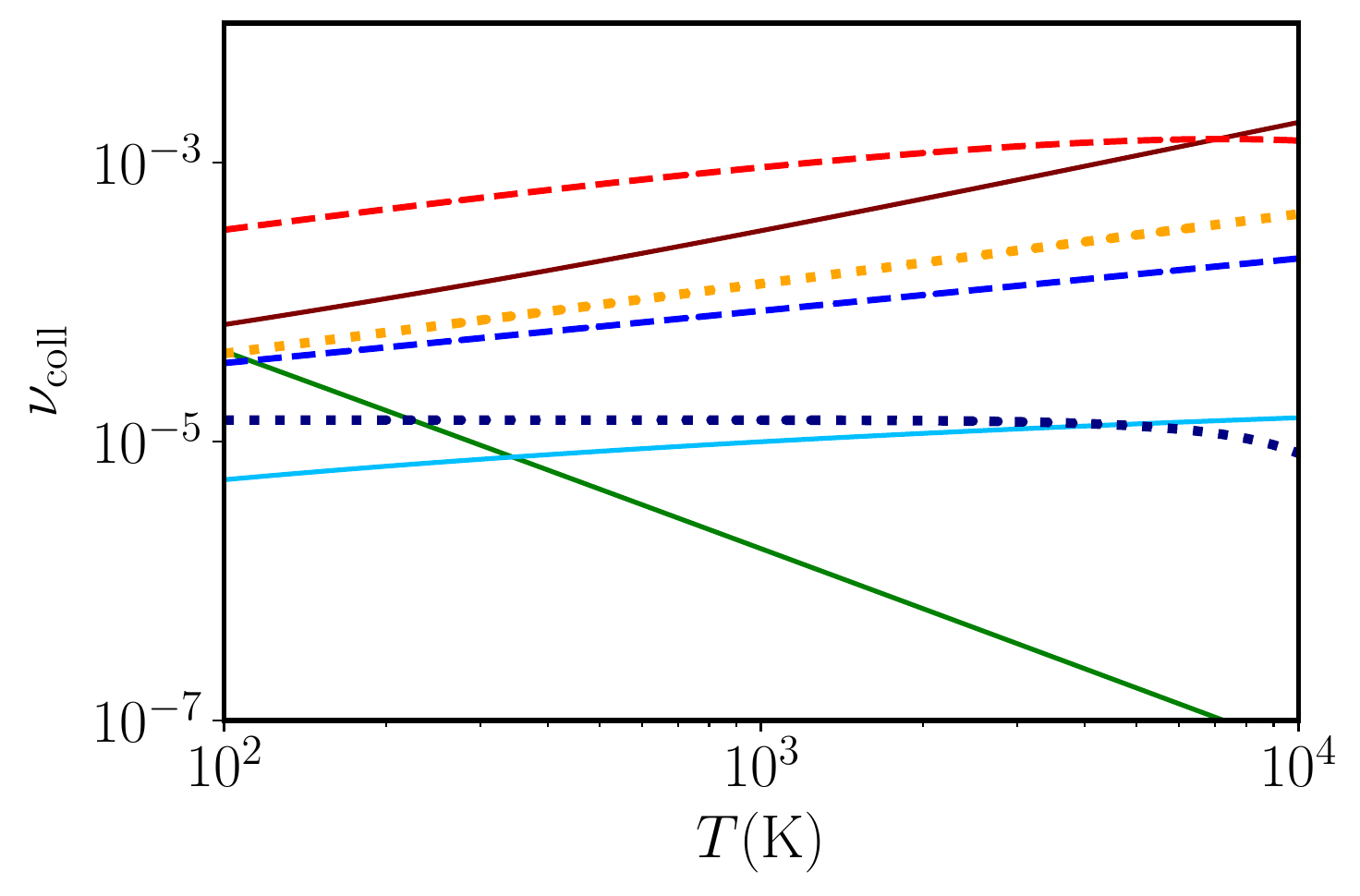}{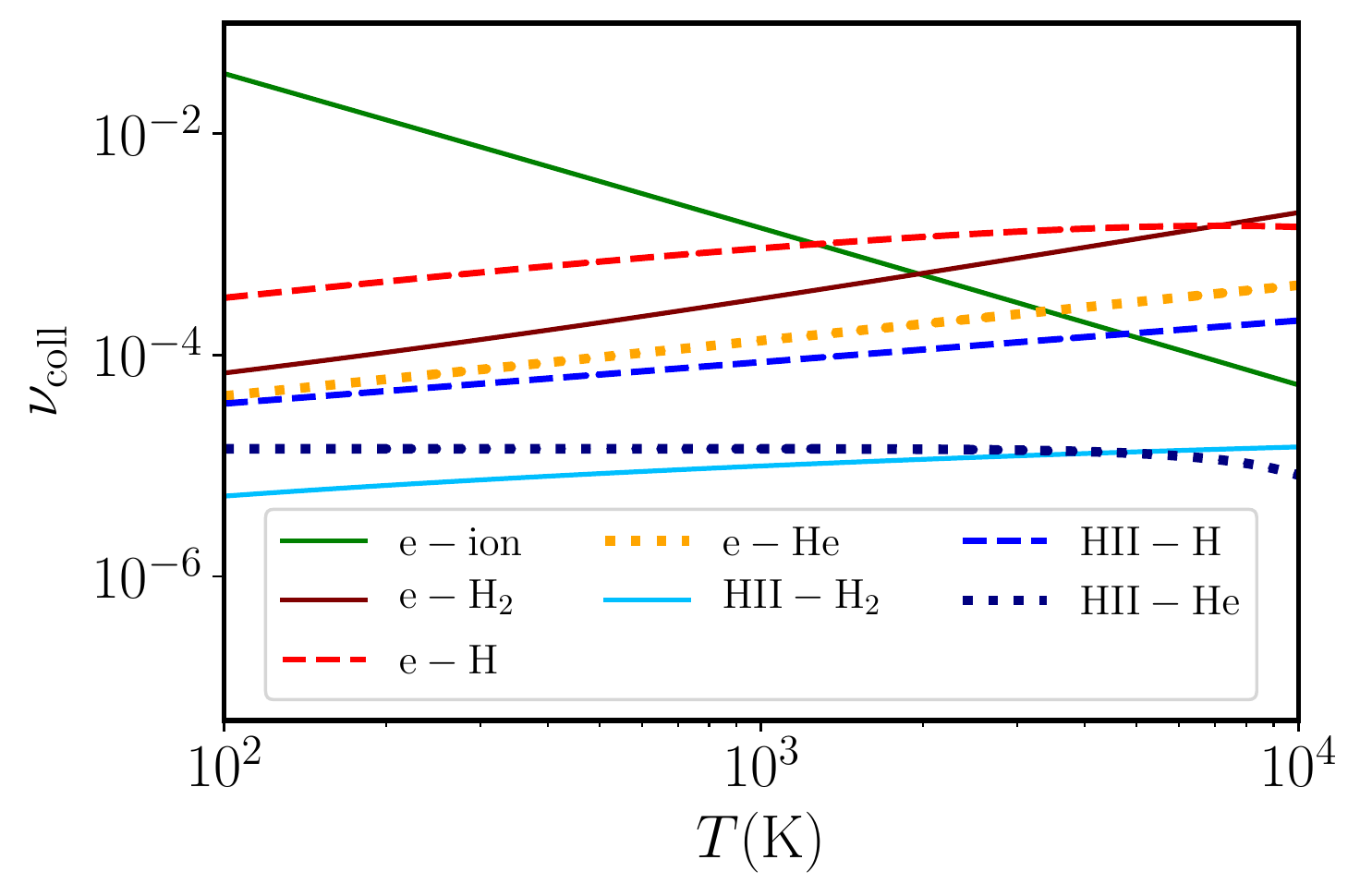}
\caption{Collision frequency between different species. Left panel shows the collision frequency at $f_i = 10^{-8}$, and the right panel is the condition when $f_i = 10^{-5}$.
\label{fig:colls}}
\end{figure}

In Fig \ref{fig:colls}, we show the collision frequency between different species for a given ionization fraction ($f_i$) and total particle number density ($n$). 
For simplicity, we set $n_e = n_i = f_{i} n$, and set the number density of each neutral species $n_s = n$. 
For $f_{i} \lesssim 10^{-8}$, corresponding to the accretion phase, electron-neutral collision dominates. Also note that for $f_{i} \gtrsim 10^{-5}$, corresponding to the collapsing phase and within the protostar, electron-ion collision becomes important.

\end{CJK}

\begin{thebibliography}{}
\providecommand{\url}[1]{\href{#1}{#1}}

\bibitem[{{Aoki} {et~al.}(2014){Aoki}, {Tominaga}, {Beers}, {Honda}, \&
  {Lee}}]{Aoki+14}
{Aoki}, W., {Tominaga}, N., {Beers}, T.~C., {Honda}, S., \& {Lee}, Y.~S. 2014,
  Science, 345, 912

\bibitem[{{Bae} {et~al.}(2016){Bae}, {Nelson}, {Hartmann}, \&
  {Richard}}]{Bae+16a}
{Bae}, J., {Nelson}, R.~P., {Hartmann}, L., \& {Richard}, S. 2016, \apj, 829,
  13

\bibitem[{{Balbus} \& {Hawley}(1991)}]{mri}
{Balbus}, S.~A., \& {Hawley}, J.~F. 1991, \apj, 376, 214

\bibitem[{{Bhat} {et~al.}(2017){Bhat}, {Ebrahimi}, {Blackman}, \&
  {Subramanian}}]{Bhat+17}
{Bhat}, P., {Ebrahimi}, F., {Blackman}, E.~G., \& {Subramanian}, K. 2017,
  \mnras, 472, 2569

\bibitem[{{Biermann}(1950)}]{Biermann+50}
{Biermann}, L. 1950, Zeitschrift Naturforschung Teil A, 5, 65

\bibitem[{{Blackman} \& {Field}(2000)}]{Blackman+00}
{Blackman}, E.~G., \& {Field}, G.~B. 2000, \apj, 534, 984

\bibitem[{{Blackman} \& {Field}(2002)}]{Blackman+02}
---. 2002, \prl, 89, 265007

\bibitem[{{Bowman} {et~al.}(2018){Bowman}, {Rogers}, {Monsalve}, {Mozdzen}, \&
  {Mahesh}}]{Bowman+18}
{Bowman}, J.~D., {Rogers}, A.~E.~E., {Monsalve}, R.~A., {Mozdzen}, T.~J., \&
  {Mahesh}, N. 2018, \nat, 555, 67

\bibitem[{{Brandenburg}(2001)}]{Brandenburg+01}
{Brandenburg}, A. 2001, \apj, 550, 824

\bibitem[{{Brandenburg}(2009)}]{Brandenburg+09}
---. 2009, \apj, 697, 1206

\bibitem[{{Brandenburg} {et~al.}(2012){Brandenburg}, {Sokoloff}, \&
  {Subramanian}}]{Brandenburg+12}
{Brandenburg}, A., {Sokoloff}, D., \& {Subramanian}, K. 2012, \ssr, 169, 123

\bibitem[{{Chandrasekhar}(1960)}]{Chandrasekhar+60}
{Chandrasekhar}, S. 1960, Proceedings of the National Academy of Science, 46,
  253

\bibitem[{{Choplin} {et~al.}(2019){Choplin}, {Tominaga}, \&
  {Ishigaki}}]{Choplin+19}
{Choplin}, A., {Tominaga}, N., \& {Ishigaki}, M.~N. 2019, arXiv e-prints,
  arXiv:1910.01366

\bibitem[{{Clark} {et~al.}(2011){Clark}, {Glover}, {Smith}, {Greif}, {Klessen},
  \& {Bromm}}]{Clark+11}
{Clark}, P.~C., {Glover}, S.~C.~O., {Smith}, R.~J., {et~al.} 2011, Science,
  331, 1040

\bibitem[{{Deng} {et~al.}(2017){Deng}, {Mayer}, \& {Meru}}]{Deng+17}
{Deng}, H., {Mayer}, L., \& {Meru}, F. 2017, \apj, 847, 43

\bibitem[{{Ezzeddine} {et~al.}(2019){Ezzeddine}, {Frebel}, {Roederer},
  {Tominaga}, {Tumlinson}, {Ishigaki}, {Nomoto}, {Placco}, \&
  {Aoki}}]{Ezzeddine+19}
{Ezzeddine}, R., {Frebel}, A., {Roederer}, I.~U., {et~al.} 2019, \apj, 876, 97

\bibitem[{{Federrath} {et~al.}(2011){Federrath}, {Sur}, {Schleicher},
  {Banerjee}, \& {Klessen}}]{Federrath+11}
{Federrath}, C., {Sur}, S., {Schleicher}, D.~R.~G., {Banerjee}, R., \&
  {Klessen}, R.~S. 2011, \apj, 731, 62

\bibitem[{{Feng} \& {Holder}(2018)}]{Feng+18}
{Feng}, C., \& {Holder}, G. 2018, \apjl, 858, L17

\bibitem[{{Glover}(2008)}]{Glover+08}
{Glover}, S. 2008, in American Institute of Physics Conference Series, Vol.
  990, First Stars III, ed. B.~W. {O'Shea} \& A.~{Heger}, 25--29

\bibitem[{{Greif}(2014)}]{Greif+14}
{Greif}, T.~H. 2014, \mnras, 444, 1566

\bibitem[{{Greif} {et~al.}(2011){Greif}, {Springel}, {White}, {Glover},
  {Clark}, {Smith}, {Klessen}, \& {Bromm}}]{Greif+11a}
{Greif}, T.~H., {Springel}, V., {White}, S.~D.~M., {et~al.} 2011, \apj, 737, 75

\bibitem[{{Harrison}(1970)}]{Harrison+70}
{Harrison}, E.~R. 1970, \mnras, 147, 279

\bibitem[{{Hennebelle} \& {Teyssier}(2008)}]{Hennebelle+08}
{Hennebelle}, P., \& {Teyssier}, R. 2008, \aap, 477, 25

\bibitem[{{Hirano} \& {Bromm}(2018)}]{Hirano+18}
{Hirano}, S., \& {Bromm}, V. 2018, \mnras, 476, 3964

\bibitem[{{Hirano} {et~al.}(2014){Hirano}, {Hosokawa}, {Yoshida}, {Umeda},
  {Omukai}, {Chiaki}, \& {Yorke}}]{Hirano+14}
{Hirano}, S., {Hosokawa}, T., {Yoshida}, N., {et~al.} 2014, \apj, 781, 60

\bibitem[{{Hosokawa} {et~al.}(2016){Hosokawa}, {Hirano}, {Kuiper}, {Yorke},
  {Omukai}, \& {Yoshida}}]{Hosokawa+16}
{Hosokawa}, T., {Hirano}, S., {Kuiper}, R., {et~al.} 2016, \apj, 824, 119

\bibitem[{{Hosokawa} {et~al.}(2011){Hosokawa}, {Omukai}, {Yoshida}, \&
  {Yorke}}]{Hosokawa+11}
{Hosokawa}, T., {Omukai}, K., {Yoshida}, N., \& {Yorke}, H.~W. 2011, Science,
  334, 1250

\bibitem[{{Ishigaki} {et~al.}(2018){Ishigaki}, {Tominaga}, {Kobayashi}, \&
  {Nomoto}}]{Ishigaki+18}
{Ishigaki}, M.~N., {Tominaga}, N., {Kobayashi}, C., \& {Nomoto}, K. 2018, \apj,
  857, 46

\bibitem[{{Iskakov} {et~al.}(2007){Iskakov}, {Schekochihin}, {Cowley},
  {McWilliams}, \& {Proctor}}]{Iskakov+07}
{Iskakov}, A.~B., {Schekochihin}, A.~A., {Cowley}, S.~C., {McWilliams}, J.~C.,
  \& {Proctor}, M.~R.~E. 2007, Physical Review Letters, 98, 208501

\bibitem[{{Latif} {et~al.}(2013){Latif}, {Schleicher}, {Schmidt}, \&
  {Niemeyer}}]{Latif+13}
{Latif}, M.~A., {Schleicher}, D.~R.~G., {Schmidt}, W., \& {Niemeyer}, J. 2013,
  \mnras, 432, 668

\bibitem[{{Lee} {et~al.}(2018){Lee}, {Hwang}, {Ching}, {Hirano}, {Lai}, {Rao},
  \& {Ho}}]{LeeCF+18}
{Lee}, C.-F., {Hwang}, H.-C., {Ching}, T.-C., {et~al.} 2018, Nature
  Communications, 9, 4636

\bibitem[{{Li} {et~al.}(2011){Li}, {Krasnopolsky}, \& {Shang}}]{LiZY+11}
{Li}, Z.-Y., {Krasnopolsky}, R., \& {Shang}, H. 2011, \apj, 738, 180

\bibitem[{{Liao} \& {Turk}(2019)}]{Liao+19}
{Liao}, W.-T., \& {Turk}, M. 2019, arXiv e-prints, arXiv:1911.00610

\bibitem[{{Lin} {et~al.}(2011){Lin}, {Krumholz}, \& {Kratter}}]{LinMK+11}
{Lin}, M.-K., {Krumholz}, M.~R., \& {Kratter}, K.~M. 2011, \mnras, 416, 580

\bibitem[{{Maeder} \& {Meynet}(2012)}]{Maeder+12}
{Maeder}, A., \& {Meynet}, G. 2012, Reviews of Modern Physics, 84, 25

\bibitem[{{Matt} \& {Pudritz}(2005)}]{Matt+05}
{Matt}, S., \& {Pudritz}, R.~E. 2005, \mnras, 356, 167

\bibitem[{{Matt} {et~al.}(2010){Matt}, {Pinz{\'o}n}, {de la Reza}, \&
  {Greene}}]{Matt+10}
{Matt}, S.~P., {Pinz{\'o}n}, G., {de la Reza}, R., \& {Greene}, T.~P. 2010,
  \apj, 714, 989

\bibitem[{{Mellon} \& {Li}(2008)}]{Mellon+08}
{Mellon}, R.~R., \& {Li}, Z.-Y. 2008, \apj, 681, 1356

\bibitem[{{Mellon} \& {Li}(2009)}]{Mellon+09}
---. 2009, \apj, 698, 922

\bibitem[{{Mignone} {et~al.}(2007){Mignone}, {Bodo}, {Massaglia}, {Matsakos},
  {Tesileanu}, {Zanni}, \& {Ferrari}}]{pluto}
{Mignone}, A., {Bodo}, G., {Massaglia}, S., {et~al.} 2007, \apjs, 170, 228

\bibitem[{{Mouschovias}(1991)}]{Mouschovias+91}
{Mouschovias}, T.~C. 1991, \apj, 373, 169

\bibitem[{{Nakauchi} {et~al.}(2019){Nakauchi}, {Omukai}, \&
  {Susa}}]{Nakauchi+19}
{Nakauchi}, D., {Omukai}, K., \& {Susa}, H. 2019, \mnras, 488, 1846

\bibitem[{{Nelson} {et~al.}(2013){Nelson}, {Gressel}, \& {Umurhan}}]{Nelson+13}
{Nelson}, R.~P., {Gressel}, O., \& {Umurhan}, O.~M. 2013, \mnras, 435, 2610

\bibitem[{{Nicholson}(1983)}]{Nicholson+83}
{Nicholson}, D.~R. 1983, {Introduction to Plasma Theory}

\bibitem[{{Pinto} \& {Galli}(2008)}]{Pinto+08}
{Pinto}, C., \& {Galli}, D. 2008, \aap, 484, 17

\bibitem[{{Pouquet} {et~al.}(1976){Pouquet}, {Frisch}, \&
  {Leorat}}]{Pouquet+76}
{Pouquet}, A., {Frisch}, U., \& {Leorat}, J. 1976, Journal of Fluid Mechanics,
  77, 321

\bibitem[{{Price} \& {Bate}(2008)}]{Price+08}
{Price}, D.~J., \& {Bate}, M.~R. 2008, \mnras, 385, 1820

\bibitem[{{Richard} {et~al.}(2016){Richard}, {Nelson}, \&
  {Umurhan}}]{Richard+16}
{Richard}, S., {Nelson}, R.~P., \& {Umurhan}, O.~M. 2016, \mnras, 456, 3571

\bibitem[{{Riols} {et~al.}(2017){Riols}, {Latter}, \&
  {Paardekooper}}]{Riols+17}
{Riols}, A., {Latter}, H., \& {Paardekooper}, S.-J. 2017, \mnras, 471, 317

\bibitem[{{Rosen} {et~al.}(2012){Rosen}, {Krumholz}, \&
  {Ramirez-Ruiz}}]{Rosen+12}
{Rosen}, A.~L., {Krumholz}, M.~R., \& {Ramirez-Ruiz}, E. 2012, \apj, 748, 97

\bibitem[{{Ruzmaikin} {et~al.}(1988){Ruzmaikin}, {Sokolov}, \&
  {Shukurov}}]{Ruzmaikin+88}
{Ruzmaikin}, A.~A., {Sokolov}, D.~D., \& {Shukurov}, A.~M. 1988, {Magnetic
  Fields of Galaxies}, Vol. 133, doi:10.1007/978-94-009-2835-0

\bibitem[{{Schekochihin} {et~al.}(2004){Schekochihin}, {Cowley}, {Taylor},
  {Maron}, \& {McWilliams}}]{Schekochihin+04_morphology}
{Schekochihin}, A.~A., {Cowley}, S.~C., {Taylor}, S.~F., {Maron}, J.~L., \&
  {McWilliams}, J.~C. 2004, \apj, 612, 276

\bibitem[{{Schive} {et~al.}(2018){Schive}, {ZuHone}, {Goldbaum}, {Turk},
  {Gaspari}, \& {Cheng}}]{gamer}
{Schive}, H.-Y., {ZuHone}, J.~A., {Goldbaum}, N.~J., {et~al.} 2018, \mnras,
  481, 4815

\bibitem[{{Schleicher} {et~al.}(2010){Schleicher}, {Banerjee}, {Sur},
  {Arshakian}, {Klessen}, {Beck}, \& {Spaans}}]{Schleicher+10}
{Schleicher}, D.~R.~G., {Banerjee}, R., {Sur}, S., {et~al.} 2010, \aap, 522,
  A115

\bibitem[{{Schober} {et~al.}(2012){Schober}, {Schleicher}, {Federrath},
  {Glover}, {Klessen}, \& {Banerjee}}]{Schober+12b}
{Schober}, J., {Schleicher}, D., {Federrath}, C., {et~al.} 2012, \apj, 754, 99

\bibitem[{{Schober} {et~al.}(2015){Schober}, {Schleicher}, {Federrath},
  {Bovino}, \& {Klessen}}]{Schober+15}
{Schober}, J., {Schleicher}, D.~R.~G., {Federrath}, C., {Bovino}, S., \&
  {Klessen}, R.~S. 2015, \pre, 92, 023010

\bibitem[{{Shakura} \& {Sunyaev}(1973)}]{Shakura+73}
{Shakura}, N.~I., \& {Sunyaev}, R.~A. 1973, \aap, 24, 337

\bibitem[{{Shukurov} {et~al.}(2006){Shukurov}, {Sokoloff}, {Subramanian}, \&
  {Brand enburg}}]{Shukurov+06}
{Shukurov}, A., {Sokoloff}, D., {Subramanian}, K., \& {Brand enburg}, A. 2006,
  \aap, 448, L33

\bibitem[{{Skinner} \& {Ostriker}(2010)}]{athena_cyl}
{Skinner}, M.~A., \& {Ostriker}, E.~C. 2010, \apjs, 188, 290

\bibitem[{{Smith} {et~al.}(2017){Smith}, {Bryan}, {Glover}, {Goldbaum}, {Turk},
  {Regan}, {Wise}, {Schive}, {Abel}, {Emerick}, {O'Shea}, {Anninos}, {Hummels},
  \& {Khochfar}}]{grackle}
{Smith}, B.~D., {Bryan}, G.~L., {Glover}, S.~C.~O., {et~al.} 2017, \mnras, 466,
  2217

\bibitem[{{Stacy} {et~al.}(2016){Stacy}, {Bromm}, \& {Lee}}]{Stacy+16}
{Stacy}, A., {Bromm}, V., \& {Lee}, A.~T. 2016, \mnras, 462, 1307

\bibitem[{{Stacy} {et~al.}(2011){Stacy}, {Bromm}, \&
  {Loeb}}]{Stacy+11_rotation}
{Stacy}, A., {Bromm}, V., \& {Loeb}, A. 2011, \mnras, 413, 543

\bibitem[{{Subramanian} \& {Brandenburg}(2006)}]{Subramanian+06}
{Subramanian}, K., \& {Brandenburg}, A. 2006, \apjl, 648, L71

\bibitem[{{Sur} {et~al.}(2010){Sur}, {Schleicher}, {Banerjee}, {Federrath}, \&
  {Klessen}}]{Sur+10}
{Sur}, S., {Schleicher}, D.~R.~G., {Banerjee}, R., {Federrath}, C., \&
  {Klessen}, R.~S. 2010, \apjl, 721, L134

\bibitem[{{Susa} {et~al.}(2014){Susa}, {Hasegawa}, \& {Tominaga}}]{Susa+14}
{Susa}, H., {Hasegawa}, K., \& {Tominaga}, N. 2014, \apj, 792, 32

\bibitem[{{Tan} \& {Blackman}(2004)}]{Tan+04_dynamo}
{Tan}, J.~C., \& {Blackman}, E.~G. 2004, \apj, 603, 401

\bibitem[{{Truelove} {et~al.}(1997){Truelove}, {Klein}, {McKee}, {Holliman},
  {Howell}, \& {Greenough}}]{Truelove+97}
{Truelove}, J.~K., {Klein}, R.~I., {McKee}, C.~F., {et~al.} 1997, \apjl, 489,
  L179

\bibitem[{{Turk} {et~al.}(2012){Turk}, {Oishi}, {Abel}, \& {Bryan}}]{Turk+12}
{Turk}, M.~J., {Oishi}, J.~S., {Abel}, T., \& {Bryan}, G.~L. 2012, \apj, 745,
  154

\bibitem[{{Turk} {et~al.}(2011){Turk}, {Smith}, {Oishi}, {Skory}, {Skillman},
  {Abel}, \& {Norman}}]{2011ApJS..192....9T}
{Turk}, M.~J., {Smith}, B.~D., {Oishi}, J.~S., {et~al.} 2011, \apjs, 192, 9

\bibitem[{{Urpin} \& {Brandenburg}(1998)}]{Urpin+98}
{Urpin}, V., \& {Brandenburg}, A. 1998, \mnras, 294, 399

\bibitem[{{Velikhov}(1959)}]{Velikhov+59}
{Velikhov}, E.~P. 1959, Sov. Phys. JETP, 36, 1398

\bibitem[{{Weibel}(1959)}]{Weibel+59}
{Weibel}, E.~S. 1959, \prl, 2, 83

\bibitem[{{Yoon} {et~al.}(2012){Yoon}, {Dierks}, \& {Langer}}]{Yoon+12}
{Yoon}, S.~C., {Dierks}, A., \& {Langer}, N. 2012, \aap, 542, A113

\end{thebibliography}
\end{document}